\documentclass[twocolumn,tighten,twocolappendix,
]{aastex701}

\usepackage[T1]{fontenc}
\usepackage{amsmath}
\usepackage{xstring}
\usepackage[print-unity-mantissa=false]{siunitx}
\newcommand{\feadme}{\texttt{FEADME}{}}
\newcommand{\numpyro}{\texttt{NumPyro}{}}
\newcommand{\jax}{\texttt{JAX}{}}

\usepackage{listings}
\usepackage{xcolor}

\definecolor{codegreen}{rgb}{0,0.6,0}
\definecolor{codegray}{rgb}{0.5,0.5,0.5}
\definecolor{codepurple}{rgb}{0.58,0,0.82}
\definecolor{backcolour}{rgb}{0.95,0.95,0.92}

\lstdefinestyle{default}{
    backgroundcolor=\color{backcolour},   
    commentstyle=\color{codegreen},
    keywordstyle=\color{magenta},
    numberstyle=\tiny\color{codegray},
    stringstyle=\color{codepurple},
    basicstyle=\ttfamily\footnotesize,
    breakatwhitespace=false,         
    breaklines=true,                 
    captionpos=b,                    
    keepspaces=true,                 
    numbers=left,                    
    numbersep=5pt,                  
    showspaces=false,                
    showstringspaces=false,
    showtabs=false,                  
    tabsize=2
}

\definecolor{eclipseStrings}{RGB}{42,0.0,255}
\definecolor{eclipseKeywords}{RGB}{127,0,85}
\colorlet{numb}{magenta!60!black}

\lstdefinelanguage{json}{
    basicstyle=\normalfont\ttfamily,
    commentstyle=\color{eclipseStrings}, 
    stringstyle=\color{eclipseKeywords}, 
    numbers=left,
    numberstyle=\scriptsize,
    stepnumber=1,
    numbersep=8pt,
    showstringspaces=false,
    breaklines=true,
    backgroundcolor=\color{backcolour}, 
    string=[s]{"}{"},
    comment=[l]{:\ "},
    morecomment=[l]{:"},
    literate=
        *{0}{{{\color{numb}0}}}{1}
         {1}{{{\color{numb}1}}}{1}
         {2}{{{\color{numb}2}}}{1}
         {3}{{{\color{numb}3}}}{1}
         {4}{{{\color{numb}4}}}{1}
         {5}{{{\color{numb}5}}}{1}
         {6}{{{\color{numb}6}}}{1}
         {7}{{{\color{numb}7}}}{1}
         {8}{{{\color{numb}8}}}{1}
         {9}{{{\color{numb}9}}}{1}
}

\newcommand{\eline}[3][a]{%
    \IfStrEqCase{#1}{%
        {a}{#2${\,}\textsc{#3}$}%
        {f}{[#2${\,}\textsc{#3}$]}%
    }[failed]%
}
\newcommand{\elinesgl}[4][a]{%
    \IfStrEqCase{#1}{%
        {a}{#2${\,}\textsc{#3}{\;} \lambda #4$}%
        {f}{[#2${\,}\textsc{#3}]{\;} \lambda #4$}%
    }[failed]%
}
\newcommand{\elinedbl}[5][a]{%
    \IfStrEqCase{#1}{%
        {a}{#2${\,}\textsc{#3}{\;} \lambda\lambda #4, #5$}%
        {f}{[#2${\,}\textsc{#3}]{\;} \lambda\lambda #4, #5$}%
    }[failed]%
}

\graphicspath{{./}{figures/}}

\begin{document}

\title{{\feadme}: Fast Elliptical Accretion Disk Modeling Engine}

\author[orcid=0000-0003-1714-7415]{Nicholas Earl}
\affiliation{Department of Astronomy, University of Illinois, 1002 W. Green St., Urbana, IL 61801, USA} 
\affiliation{Center for Astrophysical Surveys, National Center for Supercomputing Applications, Urbana, IL, 61801, USA}
\email[show]{nmearl2@illinois.edu}

\author[orcid=0000-0002-4235-7337]{K.~Decker~French}
\affiliation{Department of Astronomy, University of Illinois, 1002 W. Green St., Urbana, IL 61801, USA} 
\affiliation{Center for Astrophysical Surveys, National Center for Supercomputing Applications, Urbana, IL, 61801, USA}
\email{deckerkf@illinois.edu}

\author[orcid=0000-0001-9668-2920]{Jason T. Hinkle}
\altaffiliation{NHFP Einstein Fellow}
\affiliation{Department of Astronomy, University of Illinois, 1002 W. Green St., Urbana, IL 61801, USA} 
\affiliation{NSF-Simons AI Institute for the Sky (SkAI), 172 E. Chestnut St., Chicago, IL 60611, USA}
\email{jhinkle6@illinois.edu}

\author[orcid=0009-0004-0436-0932]{Yashasvi Moon}
\affiliation{Department of Astronomy, University of Illinois, 1002 W. Green St., Urbana, IL 61801, USA} 
\email{ymoon@illinois.edu}

\author[orcid=0009-0005-1158-1896]{Margaret~Shepherd}
\affiliation{Department of Astronomy, University of Illinois, 1002 W. Green St., Urbana, IL 61801, USA}
\email{ms169@illinois.edu}

\author[orcid=0000-0003-1535-4277]{Margaret~E.~Verrico}
\affiliation{Department of Astronomy, University of Illinois, 1002 W. Green St., Urbana, IL 61801, USA}
\affiliation{Center for Astrophysical Surveys, National Center for Supercomputing Applications, Urbana, IL, 61801, USA}
\email{verrico2@illinois.edu}

\author[orcid=0000-0001-9042-965X]{Samaresh~Mondal}
\affiliation{Department of Astronomy, University of Illinois, 1002 W. Green St., Urbana, IL 61801, USA}
\email{samaresh.astro@gmail.com}

\author[orcid=0009-0008-7581-3096,sname='Ferdinand']{Ferdinand}
\affiliation{Department of Astronomy, University of Illinois, 1002 W. Green St., Urbana, IL 61801, USA}
\email{ff10@illinois.edu}

\begin{abstract}
We present {\feadme} (Fast Elliptical Accretion Disk Modeling Engine), a GPU-accelerated Python framework for modeling broad Balmer-line emission with a relativistic elliptical accretion-disk formalism. Using \jax\ and \numpyro\ for differentiable forward modeling and Bayesian inference, {\feadme} enables reproducible analyses of large samples of disk-dominated emission-line profiles. We apply the framework to 237 double-peaked emitters (DPEs) and five tidal disruption events (TDEs) with disk-like H$\alpha$ emission, fitting three model families per spectrum and selecting the preferred model using the widely applicable information criterion (WAIC). After posterior-quality filtering, the disk-bearing active galactic nuclei (AGN) sample contains 165 sources and the TDE sample contains 27 usable epochs. Within the common elliptical-disk model, AGN occupy a broad, continuous distribution of disk geometries and kinematics summarized by five phenomenological Gaussian-mixture morphology bins, while TDE and AGN parameters overlap in radial scale, local broadening, and emissivity slope. The TDE epochs tend toward lower fitted eccentricities and broader non-disk Gaussian components. \added{Comparison with published circular-disk plus spiral-arm fits to the same AGN spectra produces substantially different distributions of inclination, radii, emissivity slope, and local broadening, demonstrating that these inferred parameters depend strongly on the treatment of profile asymmetry.} Most AGN and TDEs favor models containing a disk and an additional broad-line component, suggesting that disk emission commonly coexists with more isotropic or wind-driven gas. These results demonstrate {\feadme}'s utility for population studies \added{and the importance of a consistent asymmetry prescription when comparing disk structures across accretion environments}.
\end{abstract}


\section{Introduction}

The innermost regions of active galactic nuclei (AGN) and tidal disruption events (TDEs) host complex, multi-phase gaseous environments shaped by gravity, radiation pressure, magnetic fields, and relativistic motion \citep{shakura1973,rees1988,pariev2003,giustini2019,zhang2025}. The broad Balmer emission lines that arise in these regions are powerful diagnostics of the geometry and kinematics of gas within a few hundred to a few thousand gravitational radii of the central black hole. However, inferring the structure of these regions is challenging given that the broad-line region (BLR) is typically not spatially resolved in all but a few of the nearest and brightest AGN that have recently been probed with optical/infrared interferometry \citep{gravity2024}. The BLR encompasses gas spanning a wide range of densities and ionization parameters, with emission arising from rotating disks, disk winds, turbulent or inflowing material, and potentially eccentric or precessing streams. As a result, concrete constraints on BLR structure require models that explicitly incorporate the dynamical and relativistic effects expected in accretion flows.

Double-peaked broad Balmer profiles provide the clearest spectroscopic window into disk-dominated BLR structure. The classical interpretation, originating with the relativistic Keplerian disk models of \citet{chen1989} and further developed by \citet{eracleous1995,eracleous2003} and \citet{eracleous1998}, demonstrates that geometrically thin, optically thick disks naturally reproduce the characteristic red and blue Doppler peaks and strong asymmetries from relativistic boosting. These models predict that disk emission typically originates at radii of tens to hundreds of gravitational radii, well outside the innermost stable circular orbit. Larger samples of double-peaked emitters (DPEs) confirm these expectations and show that disk geometry, kinematics, and emissivity gradients govern the diversity of observed profile morphologies \citep{eracleous1995,strateva2003,ho2000}.

The recent systematic analysis performed by \citet{ward2024} reinforces these findings on population scales. They identify 250 DPEs among ZTF-selected variable AGN, model their Balmer-line profiles using the circular disk model of \citet{chen1989} with a non-axisymmetric component, and demonstrate that the resulting fitted parameters are consistent with disk emission being a ubiquitous component of broad-line AGN. Their results show that the inferred disk radii span orders of magnitude and that long-term profile changes in a substantial fraction of the sample are consistent with evolving disk asymmetries such as spiral arms or orbiting hotspots \citep{storchi-bergmann2003,schimoia2017}.

Disk emission, however, rarely arises from a simple, pristine Keplerian flow. A larger body of theoretical and observational work has shown that radiative transfer through disk atmospheres and outflows can significantly reshape broad-line profiles. Disk-wind models \citep{murray1996,proga2000,chajet2013} predict strong anisotropies 
and the suppression or merging of the two Doppler peaks. \citet{flohic2012} applied a detailed disk-wind radiative transfer framework to Balmer-line profiles and demonstrated that even modest winds can alter line shapes enough to obscure disk signatures that would otherwise be interpreted as classical DPEs. Similarly, \citet{bon2009} investigated the interplay between disk-like emission and the broader BLR using multi-component Gaussian, Lorentzian, and disk models, finding that many AGN with apparently single-peaked profiles may still be consistent with underlying disk emission partially masked by outflows or lower-density gas. For the objects considered here, we nevertheless treat the double-peaked or strongly asymmetric component as disk-dominated. This interpretation is motivated by the broad red and blue velocity structure expected from rotation and by previous successful disk modeling of comparable AGN and TDE spectra \citep{eracleous1995,strateva2003,hung2020,wevers2022,earl2025}. In this framework, outflows or reprocessing layers are allowed to contribute additional non-disk emission, but are not assumed to replace the ordered disk-like component.

Alternative BLR geometries such as two-phase BLR structures \citep{popovic2004}, elliptical disks \citep{eracleous1995,strateva2003}, and precessing eccentric disks \citep{eracleous1995,storchi-bergmann1997} have been invoked to explain asymmetric or transient double-peaked profiles. In a single-epoch spectrum, elliptical disk models introduce eccentricity and apsidal orientation into the velocity field, providing a compact description of asymmetric peak strengths, shoulders, and skewed wings, although circular disks with localized emissivity enhancements, including spiral arms and orbiting hotspots, can produce strongly degenerate profiles while recovering substantially different radii, inclinations, and local broadening \citep{eracleous2009,storchi-bergmann2003,schimoia2017,ward2024}. Where long-term spectroscopic monitoring is available, evolving emissivity patterns reproduce some observed line-profile variability more readily than coherent precession of a globally elliptical disk \citep{gezari2007,lewis2010,eracleous2009,tucker2021}. These models remain phenomenological, however, because the data do not uniquely identify the physical origin of the fitted perturbation, and agreement with a dynamical timescale does not by itself establish precession. The impact of this degeneracy on the inferred disk parameters is quantified for the present sample in Section~\ref{subsec:disk-caveats}.

In the context of TDEs, disk-like Balmer emission is seen in a subset of events where debris circularization proceeds efficiently, but many X-ray-bright TDEs with presumably prompt disk formation nonetheless lack obvious disk-like optical line profiles, indicating that the emergence of disk-shaped Balmer emission is not a universal outcome of disk formation. Several recent TDEs have displayed double-peaked or disk-asymmetric H$\alpha$ profiles \cite[e.g.][]{holoien2019,wevers2022,hung2020,short2020,liu2017,earl2025}, with some events showing substantial profile evolution over time. The short-lived, dynamically young disks formed in TDEs likely differ in formation and evolution timescales from long-term AGN accretion disks, making them prime subjects for investigating disk-formation physics. Radiative-transfer and disk-wind models developed for AGN \citep[e.g.,][]{flohic2012,yong2017} demonstrate that outflows, optical-depth effects, and non-axisymmetric velocity fields can substantially reshape Balmer-line profiles. Similar physical ingredients are increasingly invoked in TDEs, where emerging debris disks and reprocessing layers introduce comparable kinematic complexity \citep[e.g.,][]{hung2020,holoien2019,wevers2022,thomsen2026}.


This body of work collectively demonstrates that disk emission is likely widespread in both AGN and TDEs, even if it is only visible under certain geometric or accretion-state conditions. At the same time, it shows that robust, quantitative comparison of disk structures across these populations requires modeling frameworks capable of capturing the full range of plausible disk geometries while remaining computationally tractable for large samples. Motivated by these needs, we introduce {\feadme}, the Fast Elliptical Accretion Disk Modeling Engine, a GPU-accelerated Python package built using {\jax} and {\numpyro}. {\feadme} implements the complete relativistic elliptical disk model originally developed by \citet{eracleous1995}, with optional additional components to allow for simultaneous fitting of spectral profiles. By leveraging {\jax}'s just-in-time compilation, vectorized likelihood evaluation, and {\numpyro}'s efficient No-U-Turn Sampler (NUTS) and Stochastic Variational Inference (SVI) inference engines, {\feadme} achieves order-of-magnitude speedups compared to previous disk-modeling frameworks, making rigorous Bayesian inference feasible for large AGN and TDE samples.

In this paper, we apply {\feadme} to two complementary data sets. The first is a sample of 237 AGN DPEs identified and continuum-subtracted by \citet{ward2024}. The second is a sample of five TDEs with clear disk-like Balmer emission, chosen for their well-resolved double-peaked or highly asymmetric H$\alpha$ profiles. We adopt the same elliptical-disk formalism for both populations so that the AGN and TDE line profiles are compared within a uniform model space. Our objective is to determine whether parameters inferred from a common relativistic prescription place the two populations in distinct or overlapping regions of parameter space. The fitted eccentricity therefore measures large-scale non-axisymmetric structure in the line profile and, in AGN, may absorb emissivity patterns that are not currently implemented in {\feadme}. If TDE disks truly arise from the rapid circularization of stellar debris, they may exhibit distinct geometric or kinematic signatures compared to the long-lived, secularly evolving disks of AGN. Conversely, if their inferred disk structures are similar, it would suggest that the physics of line formation in relativistic disks is remarkably universal across transient and persistent accretion episodes.

The paper is organized as follows: in Section~\ref{sec:data}, we describe the AGN and TDE samples used in this study and summarize the spectral preparation procedures. Section~\ref{sec:framework} details the {\feadme} modeling framework, including the elliptical disk implementation, sampling strategy, and model selection approach. In Section~\ref{sec:fitting}, we present the fitting procedure and outline the three model families evaluated for each spectrum. Section~\ref{sec:results} reports the inferred disk parameters for the AGN and TDE samples, along with a clustering analysis of the AGN population. In Section~\ref{sec:discussion}, we compare the global distributions of disk parameters between AGN and TDEs and discuss their physical implications. We summarize the main conclusions in Section~\ref{sec:conclusions}.

\section{Data and Sample Selection} \label{sec:data}

\subsection{AGN Sample}

The sample of 250 double-peaked emitters (DPEs) used in this analysis was drawn from the catalog of \citet{ward2024}, who searched a parent sample of highly variable, ZTF-selected broad-line AGN for systems exhibiting double-peaked H$\alpha$ emission profiles. Their classification was based on Sloan Digital Sky Survey (SDSS) spectroscopy, an accretion disk model analysis, an automated identification pipeline, and manual verification.

\citet{ward2024} initially identified 1549 highly variable broad-line AGN from the ZTF survey \citep{bellm2019,graham2019,dekany2020} and modeled the broad H$\alpha$ emission in their SDSS spectra using a circular accretion disk model after subtracting the stellar continuum with the Penalized Pixel Fitting (pPXF) code \citep{cappellari2023}. The accretion disk model, based on the framework of \citet{chen1989}, also includes a single spiral arm component which was required to describe the asymmetry in the red and blue shoulders commonly observed in disk emitters \citep{storchi-bergmann2003}. Of the 1549 AGN, 1302 yielded successful disk model fits, while the remaining spectra were dominated by narrow-line emission, preventing reliable modeling.

The selection of the DPE sample was carried out based on the accretion disk model parameters and additional spectral features of the broad H$\alpha$ line. Their classification required that the emission feature show a clear dip or plateau between the two peaks of the broad H$\alpha$ line, a velocity offset greater than \qty{500}{\km \per \s} between the peak of the broad line and the narrow H$\alpha$ component, and a flux ratio asymmetry between the red and blue peaks exceeding 0.8. The quantitative cuts were followed up by a visual inspection to refine the selection. After these refinements, the final sample contained 250 DPEs, accounting for 19.2\% of the strongly variable broad-line AGN sample. The remaining 1052 AGN without prominent double-peaked profiles comprised the control sample. 

Following \citet{ward2024}, we use the continuum-subtracted SDSS spectra\footnote{\url{https://github.com/charlotteaward/ZTF-DPEs}} for all objects in their DPE catalog as input to our {\feadme} modeling. Although the catalog contains 250 DPEs, our final sample includes 237 AGN; thirteen spectra were excluded because the corresponding data products were missing or malformed. Removing these cases ensures that all retained spectra have consistent formatting and sufficient data quality for reliable modeling.

Although all retained sources come from a curated DPE catalog, the observed H$\alpha$ morphologies are not uniformly classical double-peaked profiles. Some spectra show shallow shoulders, strong asymmetries, or substantial blending between disk-like emission, narrow lines, and broader centrally peaked components; in other cases, a single-peaked Lorentzian-like or weakly asymmetric Gaussian-like profile can provide a plausible visual description. This diversity motivates the quantitative model-family comparison used below, which tests the disk interpretation for each catalog member instead of assuming it a priori.

\subsection{TDE Sample}



\begin{figure}
\centering
\includegraphics[width=\linewidth]{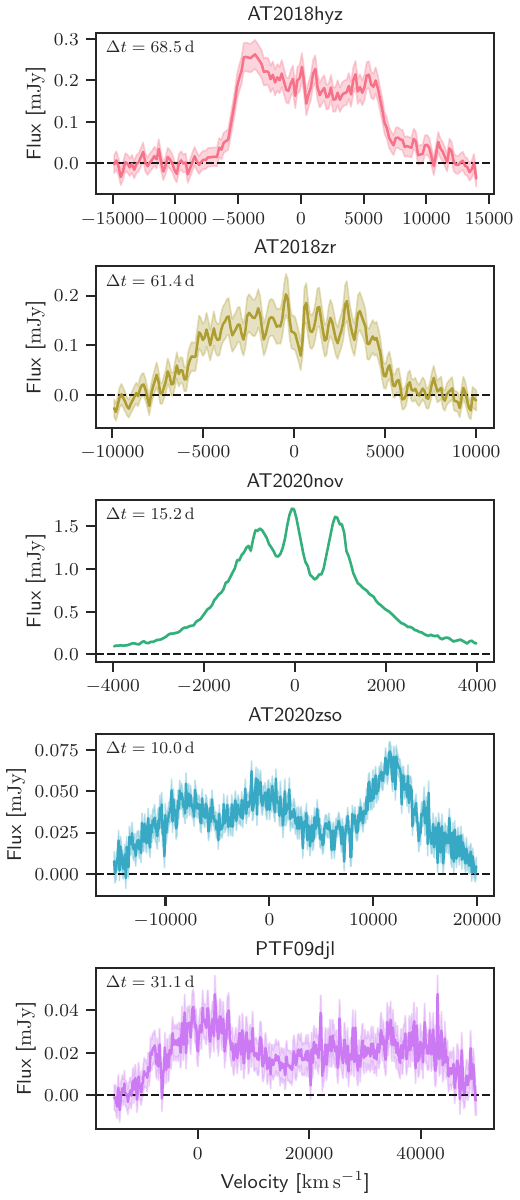}
\caption{Representative rest-frame spectra for each of the five TDEs in our sample (AT~2018hyz, AT~2018zr, AT~2020nov, AT~2020zso, and PTF09djl) showing broad, double-peaked H$\alpha$ emission consistent with disk-like kinematics. Each panel displays a single epoch selected to highlight the double-peaked structure, with the offset from peak optical brightness shown in the top-left corner. The diversity in peak separation, asymmetry, and profile shape reflects underlying differences in disk geometry, orientation, and emission structure, which we explore in detail through elliptical disk modeling in Section \ref{sec:tde-disk-structure}.
\label{fig:tde-sample}}
\end{figure}

We select five TDEs from the literature that have exhibited double-peaked emission in their Balmer line profiles: AT~2018hyz, AT~2018zr, AT~2020nov, AT~2020zso, and PTF09djl. Representative spectra for each event are shown in Figure~\ref{fig:tde-sample}. For TDEs with multiple available epochs, we retain spectra in which the broad Balmer emission shows visible double-peaked or strongly asymmetric structure, and exclude epochs where this structure has not yet emerged or has faded. This removes two late-time AT~2018hyz spectra from the final analysis set. All TDE spectra were retrieved from the WISeREP archive\footnote{\url{https://www.wiserep.org}} except for AT~2020zso, whose reduced spectra were provided through private communication.

AT~2018hyz displayed prominent double-peaked H$\alpha$ emission, with two distinct peaks separated by \qty{9320}{km \per s} \citep{hung2020, short2020}. This feature was most pronounced about \qty{51}{days} after discovery and could be well-modeled by a low eccentricity ($e \approx 0.1$) accretion disk. AT~2018zr (PS18kh) showed flat-topped and double-peaked H$\alpha$ profiles that evolved over time, though some debate exists over whether these originated from a disk or an outflow \citep{holoien2019}. AT~2020nov showed unique characteristics that blur the line between TDE and AGN accretion disks. The double-peaked emission lines, particularly prominent in the Balmer series, are well-modeled by an elliptical accretion disk characterized by moderate eccentricity ($e \approx 0.53$) and inclination ($i \approx \qty{49}{deg}$), with an unusually large outer radius of $\sim$\qty{5e4}{R_g} \citep{earl2025}. AT~2020zso demonstrated clear double-peaked profiles in multiple emission lines including H$\alpha$ and \eline{He}{ii} \citep{wevers2022}. Its H$\alpha$ profile was well-fit by a highly inclined ($i \approx \qty{85}{deg}$), highly elliptical ($e \approx 0.97$) accretion disk. PTF09djl exhibited an unusual H$\alpha$ profile with two peaks separated by \qty{3.5e4}{km \per s}, which \citet{liu2017} interpreted as emission from a highly elliptical ($e \approx 0.97$) and inclined ($i \approx \qty{88}{deg}$) disk. In the present analysis, we treat this interpretation cautiously because the available spectra have relatively low signal-to-noise compared with the other TDEs in our sample.

Other TDEs that have been fit with accretion disk models were also considered, most notably ASASSN-14li. Although ASASSN-14li has been modeled successfully with an extremely eccentric, nearly face-on disk geometry \citep{cao2018}, its observed Balmer lines are single-peaked and only moderately broad, lacking the distinct double-peaked or strongly asymmetric morphology that motivates our disk-emitter selection. We therefore exclude ASASSN-14li because its disk interpretation relies primarily on a model-dependent reconstruction and does not meet our requirement for spectroscopically evident disk-like Balmer emission.

\section{The {\feadme} Modeling Framework} \label{sec:framework}

{\feadme} (Fast Elliptical Accretion Disk Modeling Engine) is an open-source Python package for probabilistic modeling of broad emission-line profiles in astrophysical spectra. The package combines a modern Bayesian inference workflow with high-performance computation provided by {\jax} and {\numpyro}, and offers a flexible, modular interface for configuring samplers, priors, and model parameters. While the present work focuses on modeling accretion-disk emission using the elliptical disk model of \citet{eracleous1995}, {\feadme} is structured as a general tool for spectroscopic modeling, with an inference pipeline designed to accommodate additional sampling methods and modeling components as the package evolves. The version used in this analysis is archived on Zenodo \citep{earl_2026_22869620}, while ongoing development is hosted on GitHub.\footnote{\label{fn:gh-url}\url{https://github.com/nmearl/feadme}}

\subsection{Architecture and Package Design}

The {\feadme} architecture separates the physical model definition from the inference workflow, enabling a clean interface for specifying model parameters, prior distributions, likelihood functions, and sampling configurations. This design allows users to adjust or extend the inference pipeline (i.e. the choice of sampler, initialization method, etc.) without modifying the underlying physical model implementation. The package currently provides built-in support for both Stochastic Variational Inference (SVI) and Hamiltonian Monte Carlo (HMC), and is structured so that future sampling algorithms within the {\numpyro} ecosystem can be integrated with minimal overhead.

\subsection{Elliptical Accretion Disk Model}
\label{subsec:elliptical-disk}

{\feadme} implements the relativistic elliptical accretion disk model developed by \citet{eracleous1995} and adopted for large samples of disk emitters by \citet{strateva2003}. In this framework, the broad Balmer emission arises from gas in a geometrically thin, optically thick disk, with material following coplanar Keplerian ellipses. A single set of global parameters describes the disk geometry and kinematics, and the observed double-peaked line profile is obtained by integrating the locally emitted specific intensity over the entire disk surface, including relativistic Doppler and gravitational effects.

We follow the notation of \citet{strateva2003} and parameterize radii in units of the gravitational radius, $R_g$. The disk is specified by an inner and outer semimajor axis, $\xi_1$ and $\xi_2$, expressed in units of $R_g$; an eccentricity $e$ common to all orbits; an inclination angle $i$ between the disk normal and the observer's line of sight; and an azimuthal orientation angle $\phi_0$ giving the direction of apocenter in the disk plane. The line emissivity per unit area follows a power law in radius, $\epsilon(\xi) \propto \xi^{-q}$, where $q$ is a free parameter, and local line broadening is described by a Gaussian velocity dispersion $\sigma$ that encodes non-Keplerian motions (e.g., turbulence or vertical structure) in the line-emitting gas.

We introduce the dimensionless frequency shift
\begin{equation}
    X \equiv \frac{\nu}{\nu_0} - 1,
\end{equation}
where $\nu_0$ is the rest-frame line frequency. The observed, frequency-resolved line flux can then be written as
\begin{multline}
    F_X = \frac{M^2 \nu_0 \cos i}{d^2} 
    \int_{\xi_1}^{\xi_2}
    \int_{0}^{2 \pi} 
    \xi\, I_{\nu_e}(\xi, X)\, \times \\
    D^3(\xi,\phi)\,
    \Psi(\xi,\phi)\,
    d\phi\, d\xi,
    \label{eq:FX-elliptical}
\end{multline}
where $M$ is the black hole mass, $d$ is the luminosity distance, $D$ is the relativistic Doppler factor, and $\Psi(\xi,\phi)$ is a weak-field light-bending correction (see \citealt{eracleous1995,strateva2003} for the full expressions). The integral is taken over radius (expressed as the semimajor axis $\xi$) and azimuthal angle $\phi$ in the disk plane.

In the weak-field limit appropriate for $\xi \gtrsim 100\,R_g$, the light-bending factor $\Psi(\xi,\phi)$ can be expressed as
\begin{equation}
    \Psi(\xi,\phi)
    = 1 + \xi^{-1}
    \left(
        \frac{1 - \sin i \cos\phi}
             {1 + \sin i \cos\phi}
    \right),
    \label{eq:Psi}
\end{equation}
and the impact parameter $b/r$ that enters the relativistic transfer function satisfies
\begin{equation}
    \frac{b}{r} \approx
    \sqrt{1 - \sin^2 i \cos^2\phi}\, \Psi(\xi,\phi).
\end{equation}
The locally emitted specific intensity at frequency $\nu_e$ (in the emitter’s frame) is modeled as a Gaussian profile whose normalization follows the radial emissivity law:
\begin{equation}
    I_{\nu_e}(\xi, X) =
    \frac{\epsilon_0\, \xi^{-q}}{4\pi \sqrt{2\pi}\,\sigma}
    \exp\!\left[
        -\frac{(1 + X - D)^2 \nu_0^2}{2 \sigma^2 D^2}
    \right],
    \label{eq:Inue}
\end{equation}
where $\epsilon_0$ is a constant emissivity normalization and $\sigma$ is the local broadening in frequency units. 

The elliptical geometry is introduced through the relation between the semimajor axis $\xi$ and the instantaneous radius of an orbit at azimuth $\phi$:
\begin{equation}
    \xi(\phi) =
    \frac{\xi (1 + e)}{1 - e \cos(\phi - \phi_0)},
    \label{eq:xi-phi}
\end{equation}
where $e$ is the common eccentricity and $\phi_0$ specifies the orientation of apocenter. The relativistic Doppler factor $D(\xi,\phi)$ and Lorentz factor $\gamma(\xi,\phi)$ are computed following the weak-field expressions in \citet{eracleous1995} and \citet{strateva2003}, which include special relativistic Doppler shifts and boosting, transverse Doppler effects, and gravitational redshift. In the notation of those works, $D^{-1}$ is a function of $(\xi, e, \phi, \phi_0, i)$ and the impact parameter $b/r$, and $\gamma$ depends on the local orbital speed along the elliptical trajectory.

In practice, {\feadme} evaluates Equation~\ref{eq:FX-elliptical} on an observed-frame wavelength grid, using the mapping
\begin{equation}
    X = \frac{\nu}{\nu_0} - 1
      = \frac{\lambda_0}{\lambda} - 1,
\end{equation}
and transforming $F_X$ into a model line profile $F_\lambda$ on the user-specified wavelength sampling. This wavelength-space formulation allows for direct masking of contaminating features and straightforward addition of auxiliary components (e.g., narrow emission lines or an extra broad Gaussian) in the same spectral frame. It is important to note, however, that {\feadme} does not evaluate the absolute flux predicted by Equation~\ref{eq:FX-elliptical}. The multiplicative prefactor, which includes the dependence on black hole mass, distance, and the unknown absolute emissivity normalization, is factored out and replaced by a fitted integrated-flux normalization. This approach preserves the full kinematic and relativistic structure of the disk model while allowing the data to determine the overall line flux. Because the black hole masses in our AGN and TDE samples are not known with sufficient precision for absolute flux calibration, {\feadme} models the normalized line-profile shape and leaves the absolute flux scale to the fitted normalization.

The elliptical model is one plausible description of double-peaked Balmer emission, but it is not unique. Because an observed profile is a one-dimensional projection of a higher-dimensional velocity and emissivity field, distinct configurations can produce nearly identical line shapes. Within the elliptical model, $e$ and $\phi_0$ describe large-scale asymmetry in the line-emitting region. The present implementation therefore supplies a consistent, differentiable model family within which AGN and TDEs can be compared on equal terms. Extending this framework to other non-axisymmetric disk prescriptions would permit direct model comparison between competing descriptions.

The full elliptical disk model as implemented in {\feadme} is thus controlled by a set of seven free parameters,
\begin{equation}
    \left\{
    q,\,
    \xi_1,\,
    \xi_2,\,
    i,\,
    \sigma,\,
    \phi_0,\,
    e
    \right\},
\end{equation}
which together determine the radial extent, orbital shape, orientation, emissivity gradient, and local broadening of the disk. These parameters fully specify the forward model that produces the H$\alpha$ disk component used throughout this work.

The disk model is implemented in {\jax} as a fully differentiable forward model. Automatic differentiation enables efficient gradient-based sampling, while {\jax}'s vectorization and just-in-time compilation substantially reduce the computational cost of evaluating the line profile across large parameter grids or during repeated likelihood evaluations.

\subsection{{\jax} and {\numpyro} for Probabilistic Inference}

{\feadme} integrates tightly with {\numpyro} \citep{phan2019,bingham2018}, a probabilistic programming library that provides state-of-the-art Bayesian inference methods including SVI and HMC via the No-U-Turn Sampler (NUTS). Because {\numpyro} is built on {\jax}, all inference routines benefit from automatic differentiation, Accelerated Linear Algebra (XLA) compilation, and optional GPU acceleration.

The package currently supports Hamiltonian Monte Carlo (HMC) via NUTS, together with several initialization strategies designed to identify high-posterior-density basins before final sampling. In this work, we initialize NUTS using a batched, \jax-native least-squares optimizer that operates in the same differentiable model parameterization as the sampler. Operating in this common parameterization avoids reliance on a separate non-differentiable fitting model, keeps the initialization and sampling targets internally consistent, and provides stable starting positions for the expensive disk likelihood. Final posterior samples are then generated with NUTS using gradient information from the differentiable disk model. The initialization procedure adopted for the present analysis, together with the diagnostics used to assess it, is described in Section~\ref{sec:fitting}.

The sampler interface is intentionally extensible: users may configure sampler parameters, chain counts, initialization strategies, and inference diagnostics, and future versions of the package may incorporate additional inference backends.

\subsection{Command-Line Interface and Reproducible Workflows}

To facilitate large-scale modeling and reproducible analyses, {\feadme} includes a command-line interface (CLI) that allows users to run model fits, specify sampler options, define priors, and export posterior samples without writing custom Python scripts. The CLI supports both single-spectrum and batch workflow execution, making it suitable for interactive use, survey-scale pipelines, and deployment on HPC or cloud computing platforms. Example CLI usage is provided in Appendix~\ref{appendix:cli}.


\section{Model Fitting Procedure} \label{sec:fitting}

To characterize the physical properties of the broad-line emission in both the AGN and TDE samples, we fit every spectrum with the {\feadme} elliptical accretion disk framework described above. Each fit is carried out in {\numpyro} using a two-stage inference strategy: we first identify an initialization basin with a batched, \jax-native least-squares optimizer, and then draw the final posterior samples with NUTS. The initialization stage evaluates multiple structured starting points spanning the main disk-model degeneracies, including radial scale, inclination, eccentricity, emissivity slope, and the strength of any non-disk broad component. These candidates are optimized against the differentiable forward model and ranked by a posterior-based score that penalizes fragile boundary-dominated solutions, providing NUTS with a high-density starting point without changing the posterior sampled by NUTS.

Although this procedure improves robustness against a poor single start, it does not make HMC a global sampler; the basin search is intended to mitigate the multimodality inherent in decomposing a single broad-line complex into disk, narrow-line, and broad Gaussian components. Initialization therefore serves only to place NUTS within a high-posterior-density region, and all reported uncertainties and parameter correlations are derived from the final NUTS posterior. The posterior corner plots show the remaining degeneracies, most prominently those between disk area and broad Gaussian area, between inclination and radius, and between eccentricity and apocenter. These diagnostics are generated for every fit and inspected as part of the quality-tier review, with representative AGN and TDE examples shown in Appendix~\ref{appendix:posterior-diagnostics}; fits with clear chain separation, low effective sample sizes, high divergence fractions, or unresolved boundary pressure are assigned lower quality tiers or rejected according to the criteria described below.

For each object (both AGN and TDE), we construct and fit three physically motivated model families, each representing a different hypothesis for the origin of the broad H$\alpha$ emission:
\begin{enumerate}
    \item \textbf{Disk + Narrow Lines + Broad Gaussian (Full Model)}: an elliptical accretion disk component, Gaussian narrow-line components for the H$\alpha$ complex (H$\alpha$, \elinedbl[f]{N}{ii}{6548}{6584}, and \elinedbl[f]{S}{ii}{6716}{6731} where present), and an additional broad Gaussian centered on H$\alpha$ to capture non-disk high-velocity emission such as BLR contributions or winds.
    \item \textbf{Disk + Narrow Lines Only (No-BLR Model)}: the same elliptical disk and narrow-line components as above, but without the broad Gaussian. This model tests whether a disk plus narrow emission alone can reproduce the observed profile.
    \item \textbf{Broad Gaussian + Narrow Lines Only (No-Disk Model)}: a broad H$\alpha$ Gaussian together with the narrow-line components, but no disk. This provides a BLR-like baseline model against which we can assess whether disk kinematics are statistically required.
\end{enumerate}

For the AGN sample, all narrow-line components are modeled directly. Each narrow line is modeled as a single Gaussian, and the narrow H$\alpha$, \elinedbl[f]{N}{ii}{6548}{6584}, and \elinedbl[f]{S}{ii}{6716}{6731} components share a common velocity width. For the TDE sample, host-galaxy narrow-line emission was removed as part of the respective prior analyses and published reductions, whenever a high-quality host reference spectrum was available. This host subtraction was possible for all TDEs in our sample except AT~2020nov, whose narrow lines are therefore retained and modeled explicitly in the same manner as for the AGN. Apart from this external preprocessing step, all TDE spectra are fit with the same three model families as the AGN.

Each model family is sampled with two independent NUTS chains. To determine which of the three model families best describes each spectrum, we perform model comparison using the widely applicable information criterion (WAIC; an estimate of out-of-sample predictive accuracy computed from the pointwise log likelihood) as implemented in \texttt{ArviZ}. For each object we compute WAIC for the three posterior ensembles and adopt the model with the highest expected predictive accuracy as the preferred physical description. Across the \num{237} AGN in our sample, the raw WAIC selection favors the Full Model for \num{146} sources (61.6\%), the No-BLR Model for \num{68} sources (28.7\%), and the No-Disk Model for \num{23} sources (9.7\%). The disk-bearing subset used for disk-parameter and clustering analyses consists of sources whose preferred model includes an elliptical disk component, i.e., the Full Model or No-BLR Model. After quality-tier filtering, \num{188} AGN remain in the population analysis sample: \num{107} (56.9\%) prefer the Full Model, \num{58} (30.9\%) prefer the No-BLR Model, and \num{23} (12.2\%) prefer the No-Disk Model, leaving \num{165} quality-filtered disk-bearing AGN. The TDE sample shows a stronger preference for disk-bearing models, with \num{26} of the \num{29} modeled epochs preferring the Full Model and the remaining \num{3} preferring the No-BLR Model. After the epoch selection described in Section~\ref{sec:data}, the final TDE analysis sample contains \num{27} epochs, of which \num{25} (92.6\%) favor the Full Model and \num{2} (7.4\%) favor the No-BLR Model. In all subsequent analyses we use the posterior samples associated with each object's preferred WAIC-selected model family instead of enforcing a single model choice across the full sample.

We monitor the Gelman--Rubin convergence diagnostic ($\hat{R}$), effective sample sizes, divergence fractions, and posterior mass near parameter boundaries. Fits are assigned quality tiers after model selection: ``gold'' fits have clean posteriors and no significant boundary flags, ``silver'' fits have acceptable posteriors but elevated divergences or boundary pressure, ``bronze'' fits have marginal but interpretable posterior diagnostics, and ``reject'' fits have severe unresolved convergence failures. Unless stated otherwise, the population-level analysis uses the gold, silver, and bronze tiers. The quality-tier filtering removes \num{49} AGN classified as reject, leaving \num{188} AGN in the gold, silver, or bronze tiers (\num{41}, \num{110}, and \num{37} objects, respectively). Of the \num{165} disk-bearing AGN used for the disk-parameter and clustering analyses, \num{18} are gold, \num{110} are silver, and \num{37} are bronze. All \num{27} TDE epochs in the final analysis subset are classified as silver.

\section{Results} \label{sec:results}

\subsection{Disk Parameters Across AGN Sample}

\begin{figure*}
\centering
\includegraphics[width=\linewidth]{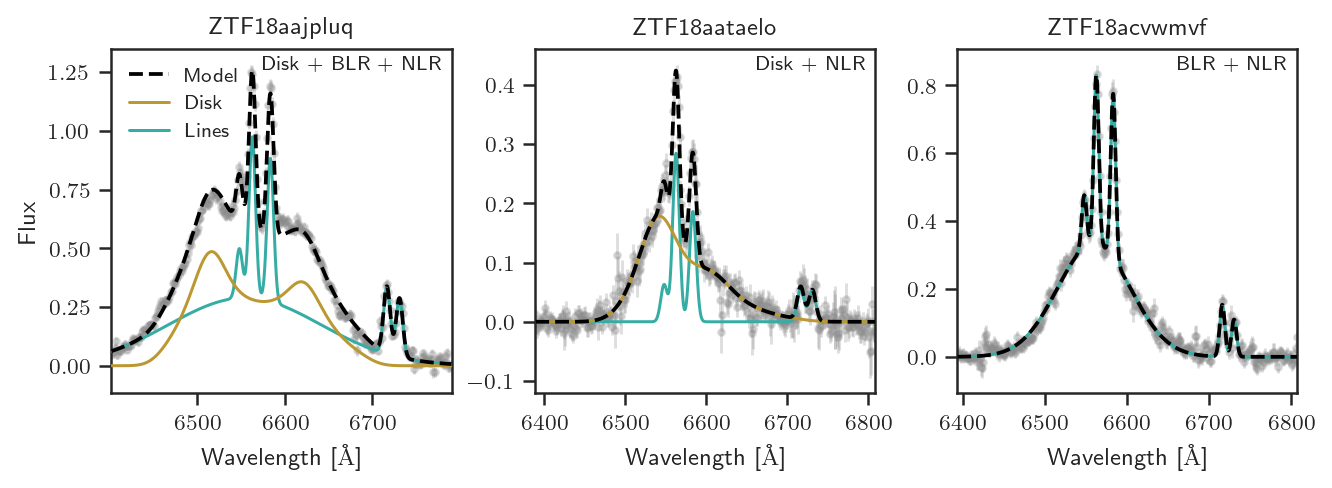}
\caption{Representative examples of the three model families used to fit the AGN spectra in this work. Each panel shows an observed AGN spectrum (grey), together with the corresponding best-fit model (black dashed line). The contribution from the elliptical accretion disk is shown in gold, while all non-disk line-emission components (narrow lines and the broad Gaussian) are shown in cyan. These examples illustrate the qualitative behavior of the disk+BLR, no BLR, and no disk model families (see Section \ref{sec:fitting}) and highlight the different ways in which disk and non-disk components combine to reproduce the observed H$\alpha$ line profiles.
\label{fig:agn-fits}}
\end{figure*}

\begin{figure*}
\centering
\includegraphics[width=\linewidth]{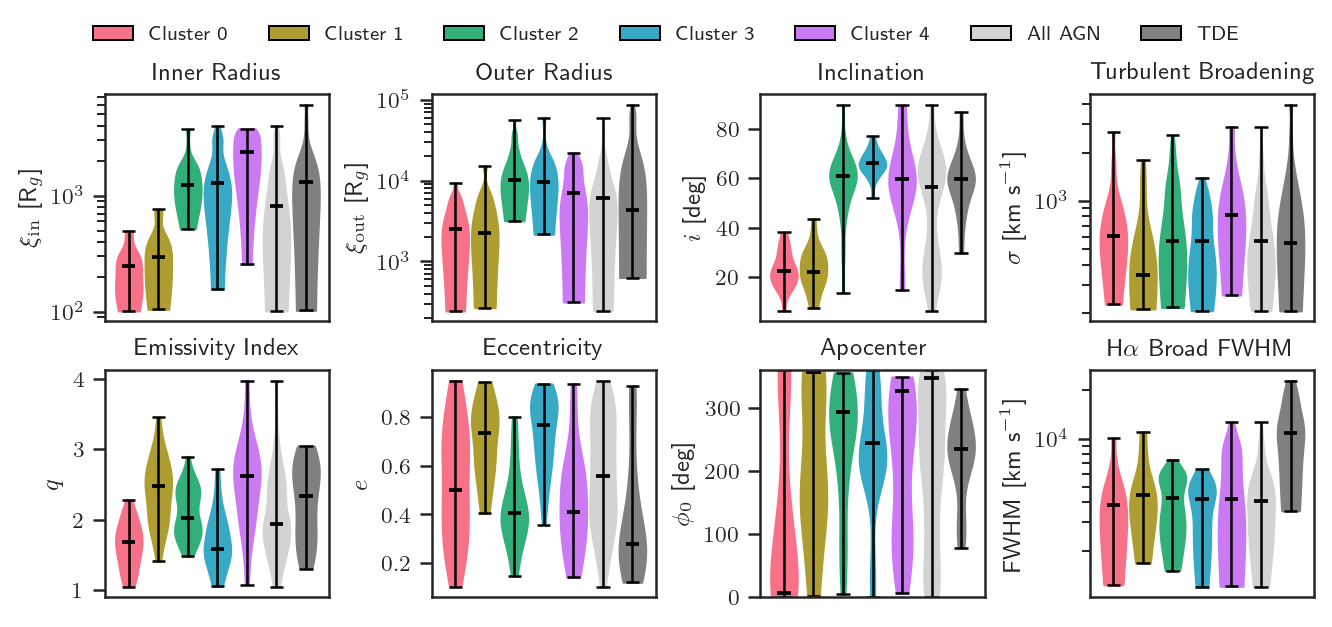}
\caption{Distribution of elliptical disk model parameters across the five AGN morphology bins identified with Gaussian-mixture binning (see Section \ref{sec:clustering-analysis}), as well as for the full disk-bearing AGN sample (light grey) and the full TDE sample (dark grey). The eight panels show the seven elliptical disk parameters (inclination ($i$), emissivity index ($q$), turbulent broadening ($\sigma$), inner radius ($\xi_1$), outer radius ($\xi_2$), apocenter angle ($\phi_0$), and eccentricity ($e$)) and the FWHM of the broad Gaussian component included when preferred by the fit. Widths indicate the relative density of the posterior-summary values for the objects or epochs in each group. The central horizontal black bar marks the group median, the vertical black line spans the range of values, and the horizontal caps mark the minimum and maximum values. The TDE sample overlaps the AGN morphology bins in most parameters and shows systematically lower eccentricities and larger non-disk broad-line widths.
\label{fig:violin-parameter-compare}}
\end{figure*}

Across the AGN sample, the elliptical disk fits favor moderately inclined, outer-disk Balmer-emitting regions with frequent eccentric structure and a wide range of emissivity slopes and local broadening values. Representative examples of these fits, illustrating the decomposition into disk and non-disk components, are shown in Figure~\ref{fig:agn-fits}. The combined parameter distributions for all disk-bearing AGN are summarized in the final row of Table~\ref{tab:cluster_params_compact} and shown as grey violins in Figure~\ref{fig:violin-parameter-compare}. The typical disk is moderately inclined, with a median inclination of $i = \qty{56.3(35.9:10.5)}{deg}$; the uncertainties quoted in this section represent the 68th percentile ranges of the fitted parameter distributions. The inner and outer radii cluster around $\xi_1 = \qty{821(609:1156)}{R_g}$ and $\xi_2 = \qty{6047(4162:9190)}{R_g}$, respectively, with long tails toward larger radii. These scales are consistent with a Balmer-emitting layer in the outer disk, far from the innermost stable circular orbit. \added{Because these values are conditional on the elliptical-disk prescription, we compare the shared parameters with the circular-disk plus spiral-arm fits of \citet{ward2024} in Section~\ref{sec:discussion}.}

Within the elliptical-disk parameterization, the fitted orbits are typically non-circular, with a median eccentricity of $e = \num{0.56(0.24:0.26)}$ indicating that most sources require appreciable non-axisymmetric structure to account for their asymmetric shoulders and unequal peak heights. \added{However, circular-disk models can reproduce similar asymmetries using spiral arms or related emissivity perturbations \citep{eracleous2009,ward2024}, and this alternative prescription produces systematically different radii, inclinations, and local broadening, as quantified in Section~\ref{subsec:disk-caveats}. We therefore interpret $e$ as a measure of the large-scale departure from circular symmetry within the elliptical model.} The apocenter orientation angle, $\phi_0$, spans nearly the full \qtyrange{0}{360}{deg} range, with a median of $\phi_0 = \qty{227.4(183.4:121.2)}{deg}$, consistent with no preferred apsidal alignment across the sample.

The radial emissivity and local broadening parameters further highlight the physical diversity of the AGN disks. The emissivity index has a median value of $q = \num{1.94(0.44:0.64)}$, with the wide range indicating that some systems are dominated by emission from the inner disk, while others exhibit relatively shallow emissivity gradients in which the outer regions contribute significantly to the line flux. This spread in $q$ is qualitatively consistent with expectations from radiative transfer calculations in disk atmospheres and winds, where the balance between local heating, optical depth, and ionization can shift the dominant line-emitting radius over time \citep[e.g.][]{flohic2012,yong2017}. The local Gaussian broadening parameter has a median of $\sigma = \qty{559(259:639)}{\km \per \s}$ and spans a broad range across the sample. In the context of the phenomenological disk model, this parameter captures unresolved local velocity dispersion beyond the ordered elliptical disk velocity field, which may include turbulence, vertical structure, radiative-transfer broadening, or small-scale flows within the emitting layer. Similar combinations of ordered rotation and disordered motions have been inferred from detailed modeling of individual broad-line AGN, particularly in systems where disk-wind components interact with the dense disk atmosphere \citep[e.g.][]{bon2009,kollatschny2013}.

Finally, the broad Gaussian component that is included when preferred by the WAIC model comparison has a characteristic FWHM of $\mathrm{FWHM}_\mathrm{broad} = \qty{4076(1980:2319)}{\km \per \s}$. This velocity scale is comparable to classical BLR widths and indicates that many AGN require an additional, roughly centrally peaked broad component in addition to the thin elliptical disk in order to reproduce their H$\alpha$ profiles. Among the quality-filtered disk-bearing AGN, \num{107}/\num{165} (64.8\%) prefer the model that includes this broad Gaussian, substantially higher than the \num{11}/\num{71} (15\%) reported by \citet{ward2025bass} for X-ray selected BASS DPEs. Some of this difference is likely methodological, since the BASS analysis fits circular disks with spiral arms, which can absorb centrally peaked or asymmetric emission that our elliptical-disk family assigns instead to the broad Gaussian. The two samples also differ in selection, with the BASS sources having been drawn from a hard X-ray survey, and our sample selected using optical variability. Taken together, the global parameter distributions show that AGN DPEs occupy a continuous and extended region of disk-parameter space, with no single canonical configuration dominating the sample. The disks are typically extended, moderately inclined, and frequently non-axisymmetric, with a wide range of emissivity slopes and local broadening, and they are often accompanied by a broader BLR-like or wind-like component. This picture is in line with earlier interpretations of complex broad-line morphologies in AGN as the combined outcome of disk rotation, non-axisymmetric structure, and outflows \citep[e.g.][]{eracleous1998,bon2009,flohic2012,ward2024,ward2025bass}, and it provides the baseline against which we compare both the cluster-specific archetypes and the TDE disk parameters in the following sections.

\begin{deluxetable*}{lccccccccc}
\tablecaption{Fitted AGN disk parameters by GMM morphology bin.\label{tab:cluster_params_compact}}
\tablehead{
\colhead{ID} & \colhead{$n$} & \colhead{$e$} & \colhead{$i$} & \colhead{$\xi_1$} & \colhead{$\xi_2$} & \colhead{$q$} & \colhead{$\sigma$} & \colhead{$\phi_0$} & \colhead{$\mathrm{FWHM}$} \\
\colhead{} & \colhead{} & \colhead{} & \colhead{[$\mathrm{deg}$]} & \colhead{[$\mathrm{R_g}$]} & \colhead{[$\mathrm{R_g}$]} & \colhead{} & \colhead{[$\unit{\km \per \s}$]} & \colhead{[$\mathrm{deg}]$} & \colhead{[$\unit{\km \per \s}$]}
}
\startdata
C0 & 31 & $0.50^{+0.26}_{-0.24}$ & $22.5^{+8.1}_{-5.6}$ & $247^{+78}_{-118}$ & $2514^{+2970}_{-1875}$ & $1.69^{+0.29}_{-0.43}$ & $601^{+360}_{-192}$ & $42.3^{+308.8}_{-38.9}$ & $3847^{+1318}_{-2248}$ \\
C1 & 31 & $0.74^{+0.11}_{-0.19}$ & $22.1^{+7.3}_{-5.9}$ & $296^{+137}_{-173}$ & $2264^{+3937}_{-1551}$ & $2.48^{+0.37}_{-0.54}$ & $343^{+762}_{-124}$ & $183.3^{+167.3}_{-98.6}$ & $4455^{+3420}_{-1957}$ \\
C2 & 44 & $0.41^{+0.14}_{-0.11}$ & $60.9^{+5.2}_{-11.8}$ & $1236^{+598}_{-454}$ & $10097^{+13019}_{-4585}$ & $2.03^{+0.42}_{-0.30}$ & $564^{+847}_{-229}$ & $273.8^{+75.0}_{-117.3}$ & $4244^{+1673}_{-1990}$ \\
C3 & 37 & $0.77^{+0.12}_{-0.18}$ & $66.1^{+4.3}_{-3.7}$ & $1294^{+1684}_{-566}$ & $9483^{+8555}_{-5142}$ & $1.59^{+0.41}_{-0.24}$ & $559^{+468}_{-298}$ & $227.7^{+103.5}_{-27.5}$ & $4205^{+1082}_{-2055}$ \\
C4 & 22 & $0.41^{+0.20}_{-0.15}$ & $59.5^{+13.0}_{-7.4}$ & $2397^{+993}_{-1256}$ & $7105^{+8267}_{-5015}$ & $2.62^{+0.34}_{-0.66}$ & $816^{+770}_{-408}$ & $233.9^{+87.9}_{-153.8}$ & $4221^{+4276}_{-2932}$ \\ \hline
All & 165 & $0.56^{+0.26}_{-0.24}$ & $56.3^{+10.5}_{-35.9}$ & $821^{+1156}_{-609}$ & $6047^{+9190}_{-4162}$ & $1.94^{+0.64}_{-0.44}$ & $559^{+639}_{-259}$ & $227.4^{+121.2}_{-183.4}$ & $4076^{+2319}_{-1980}$ \\
\enddata
\tablecomments{Values are medians and 16th--84th percentile ranges of the source-level posterior medians under the {\feadme} elliptical-disk prescription. \added{The fitted eccentricity and all shared disk parameters are conditional on this representation of profile asymmetry. Figure~\ref{fig:tde-agn-parameter-distributions} and Section~\ref{sec:discussion} compare the shared parameters with circular-disk plus spiral-arm fits.}}
\end{deluxetable*}


\subsection{Clustering Analysis} \label{sec:clustering-analysis}

To investigate how the disk-bearing AGN are distributed in parameter space, we perform a morphology analysis using the posterior medians of the WAIC-selected best-fit disk model parameters for each source. This analysis tests whether the AGN occupy strongly clustered regions of disk-parameter space and, regardless of the clustering strength, defines a set of representative AGN disk configurations against which the TDE parameter distributions can be compared.

Before constructing the morphology bins, we transform the disk parameters into a scaled feature space. We use $\log_{10}\xi_1$, $\log_{10}(\xi_2/\xi_1)$, $\cos i$, $\log_{10}\sigma$, $q$, $e$, and the circular coordinates $\cos\phi_0$ and $\sin\phi_0$. The apocenter coordinates are downweighted relative to the other parameters to prevent disk orientation from dominating the Euclidean distances in cases where the overall disk geometry is otherwise similar. We then apply the \texttt{RobustScaler} transformation implemented in \texttt{scikit-learn}, centering each feature on its median and scaling it by its interquartile range.

We assess the degree of intrinsic clustering using the Hopkins statistic \citep{hopkins1954}, which measures whether points in a feature space are more clustered than a spatially random sample. Values near $H=0.5$ indicate little departure from randomness, values approaching $H=1$ indicate strong clustering, and values below $H=0.5$ indicate a more regular or evenly spaced distribution. For the AGN disk-parameter vectors, we obtain $H = 0.609$, indicating that the AGN occupy a continuous disk-parameter manifold with only a mild clustering tendency and no cleanly separated subpopulations. We nevertheless use a fixed five-component Gaussian mixture model (GMM) to divide this manifold into phenomenological bins, for which five components provide a compromise between interpretability, population size per bin, visual separation in the projected feature space, and distinct median profile morphologies for comparison with the TDE sample. We do not interpret these bins as discrete physical classes.

\begin{figure}
\centering
\includegraphics[width=\linewidth]{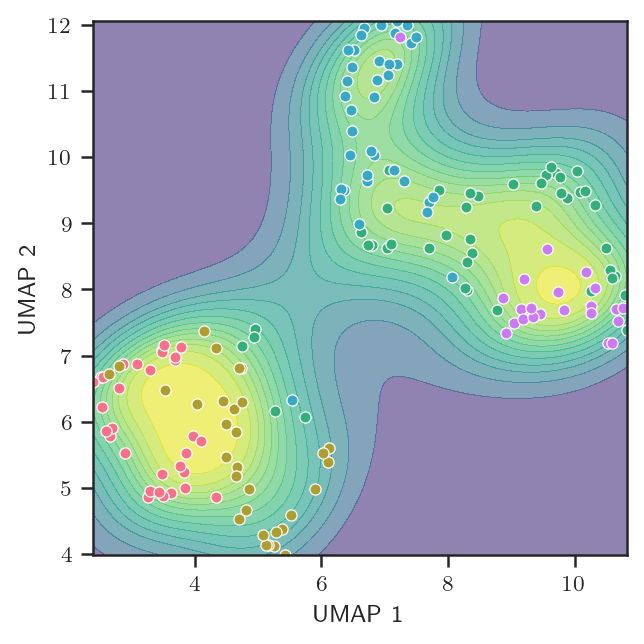}
\caption{Two-dimensional UMAP projection of the scaled AGN disk-parameter feature space. Colors indicate the five Gaussian-mixture morphology bins used in this work; the GMM itself is fit in the full scaled feature space, and UMAP is used only for visualization. The projection highlights the continuity of the AGN disk-parameter manifold and the approximate separation of the phenomenological bins used for comparison with the TDE sample.
\label{fig:agn-umap-clusters}}
\end{figure}

\begin{figure}
\centering
\includegraphics[width=\linewidth]{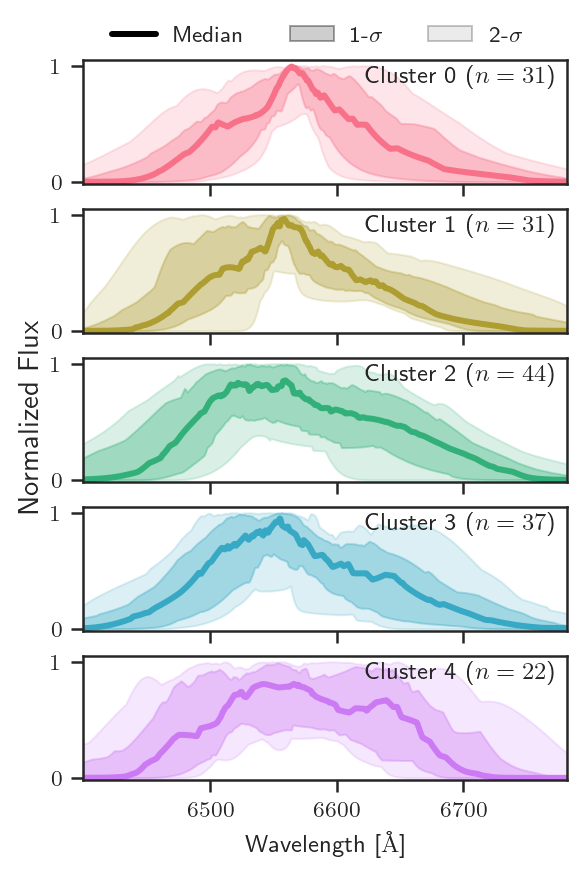}
\caption{Median rest-frame H$\alpha$ reconstructed disk-component profiles for each of the five AGN Gaussian-mixture morphology bins. The figure shows only the fitted elliptical-disk component, not the observed data or the total model including narrow lines and the additional broad Gaussian. Solid lines represent the median normalized disk profile within each bin, while the shaded regions denote the 16th--84th percentile range (dark shading) and the 5th--95th percentile range (light shading) at each wavelength. Individual reconstructed disk profiles were normalized and interpolated onto a common grid prior to stacking. The profiles illustrate the diversity in disk-component line shape and structure across the bins, with some groups showing compact centrally concentrated disk profiles and others exhibiting broader, more extended, or more asymmetric disk emission.
\label{fig:agn-cluster-medians}}
\end{figure}

The GMM assigns \num{165} disk-bearing AGN to five morphology bins with sizes \num{31}, \num{31}, \num{44}, \num{37}, and \num{22}. The resulting bins are shown in Figure~\ref{fig:agn-umap-clusters}, where the same scaled feature space is projected into two dimensions with UMAP for visualization only. The median reconstructed disk-component profiles in Figure~\ref{fig:agn-cluster-medians} illustrate the range of disk-only line shapes represented by these bins, from compact low-inclination configurations to more extended, moderately inclined disks with different emissivity slopes and eccentricities. Throughout the rest of the paper, we treat these clusters as phenomenological AGN morphology bins, not as sharply bounded physical classes.


\subsection{TDE Disk Structure} \label{sec:tde-disk-structure}

\begin{figure*}
\centering
\includegraphics[width=\linewidth]{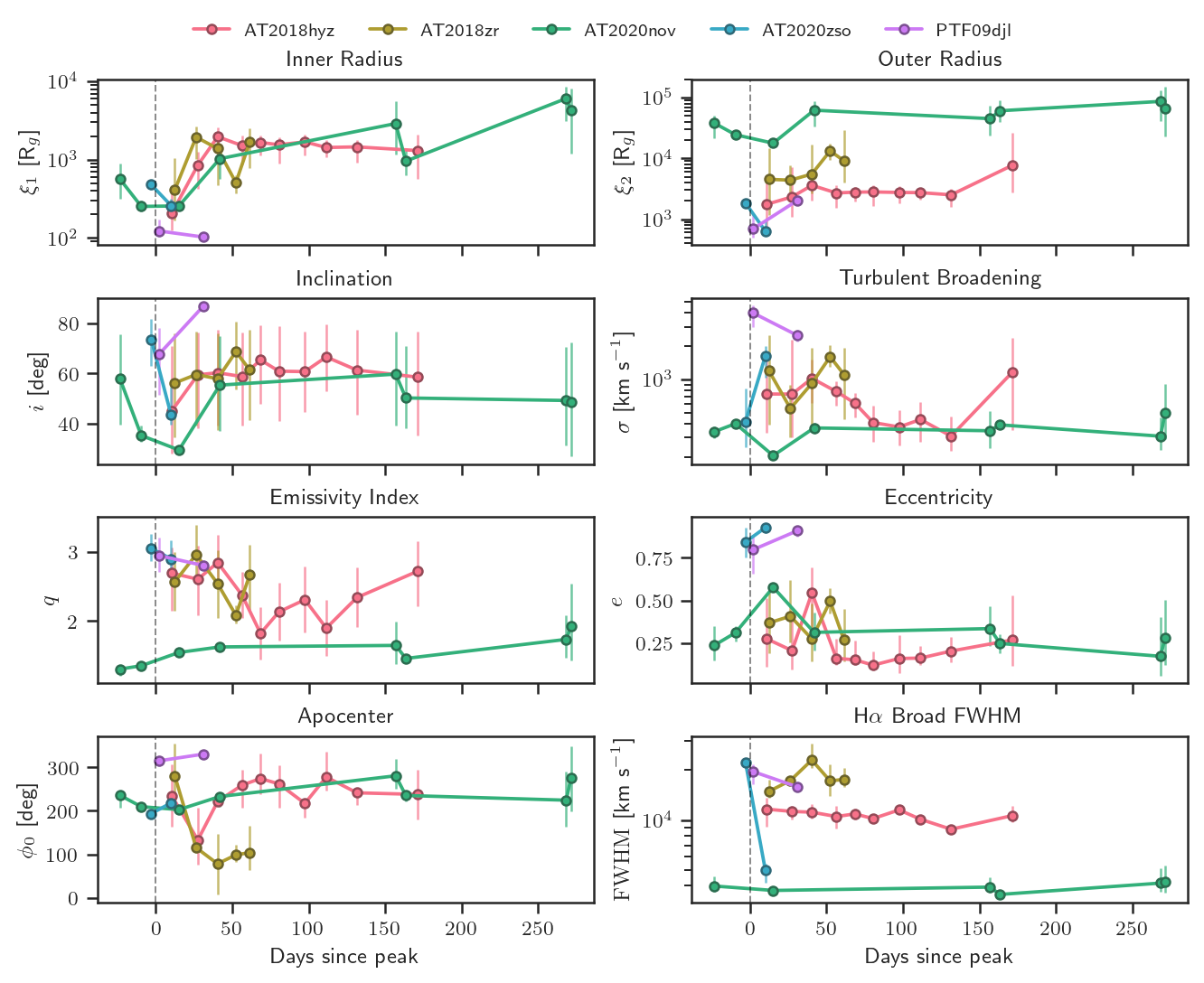}
\caption{Time evolution of the fitted elliptical disk model parameters for the five TDEs in our sample. Each panel shows one of the eight parameters (the seven disk parameters along with the FWHM of the broad Gaussian component when included). Time is normalized individually for each TDE, with $t=0$ corresponding to the epoch of maximum observed brightness in the optical light curves. Points represent the posterior medians at each epoch, with error bars indicating the 16th–84th percentile credible intervals. The evolving parameter values reflect changes in the line-emitting region and disk structure over time.
\label{fig:tde-parameter-evolution}}
\end{figure*}

The TDEs in our sample exhibit substantial diversity in their inferred disk geometries and kinematics, reflecting the range of conditions under which debris circularizes and begins emitting Balmer-line radiation. The full set of fitted parameters for all epochs is provided in Table~\ref{tab:tde_params_compact}. Their aggregate distribution relative to the AGN morphology bins is shown in Figure~\ref{fig:violin-parameter-compare}, examples of the temporal evolution for each event are shown in Figure~\ref{fig:tde-parameter-evolution}, and the individual-event contributions to the aggregate distributions are shown in Figure~\ref{fig:tde-agn-parameter-distributions}.

A defining feature of the TDE disks is the wide range of eccentricities recovered across epochs. The aggregate TDE distribution has median $e = \num{0.28(0.11:0.29)}$, lower than the AGN median but not uniformly circular. Several epochs in AT~2020zso and PTF09djl remain highly eccentric, while many AT~2018hyz epochs favor lower eccentricities. This behavior suggests that in many epochs the line-emitting debris has already undergone substantial circularization by the time a resolved double-peaked Balmer profile is visible, although several events still retain appreciable eccentricity \citep{hung2020,krolik2016}.

The inferred inclinations cluster around $i = \qty{59.5(10.7:6.9)}{deg}$, indicating that the Balmer emission is generally produced in moderately inclined viewing geometries. These inclinations result in the broad, double-peaked, or asymmetric structures observed in the line profiles of our TDE sample, all of which require appreciable projected rotational velocities to reproduce their observed peak separations.

The radial scales of the line-emitting regions span a broad range. The median inner radius across all TDE epochs is $\xi_1 = \qty{1308(1051:597)}{R_g}$, while the median outer radius is $\xi_2 = \qty{4385(2296:39842)}{R_g}$. The very broad upper tail is dominated by AT~2020nov, whose epochs frequently favor outer radii of several $10^4\,R_g$, consistent with models suggesting that its Balmer emission arose from a larger, pre-existing disk illuminated during the flare \citep{earl2025}. This places AT~2020nov closer to the extended-radius tail of the AGN disk-emitter population than to the more compact TDE disks in our sample. In contrast, AT~2018hyz and AT~2020zso often exhibit more compact disks with line-emitting regions confined to a few thousand gravitational radii.

The radial emissivity gradients likewise reflect varied disk conditions. The aggregate TDE median is $q = \num{2.34(0.71:0.50)}$, with many epochs favoring steep profiles in which the inner radii dominate the Balmer emission. Such steep emissivities are consistent with radiative-transfer expectations for dynamically young disks, where heating and ionization are concentrated near the circularization radius \citep{flohic2012}. The local turbulent broadening parameter has median $\sigma = \qty{549(200:643)}{\km \per \s}$, but individual epochs reach well above \qty{1000}{\km \per \s}. This variation may reflect differences in the vertical structure and turbulence of the line-emitting layer, as suggested by studies that decompose broad-line widths into rotational and turbulent components and relate the latter to the disk scale height \citep[e.g.][]{kollatschny2012,kollatschny2013}, as well as additional broadening from shocks and highly disturbed flows during debris circularization and disk assembly in TDEs \citep[e.g.][]{shiokawa2015,piran2015,holoien2019}.

In addition to the elliptical disk, most TDE spectra require a broad Gaussian component with characteristic $\mathrm{FWHM} = \qty{10880(6759:6486)}{\km \per \s}$. This necessity is consistent with previous interpretations of TDE Balmer-line formation, where disk emission is accompanied by contributions from reprocessing layers, outflowing gas, or high-velocity debris streams \citep[e.g.][]{holoien2019,short2020}. AT~2020nov is again an exception, since the broad Gaussian has a typical FWHM of only $\sim\qty{4000}{\km \per \s}$ when preferred, comparable to the AGN broad-component median and substantially narrower than the TDE median. Together with its large inferred outer radii, this supports the interpretation that AT~2020nov occupies a transitional regime in which a TDE flare illuminates or perturbs an AGN-like pre-existing disk. The integrated flux and width of the broad component often evolve with time in the remaining TDEs, paralleling changes in continuum luminosity and debris dynamics.

AT~2020nov may contain a pre-existing AGN-like disk that was illuminated during the TDE. We therefore also examine the aggregate TDE parameters with its epochs removed. The median eccentricity and inclination change little, from $e=\num{0.28}$ and $i=\qty{59.5}{deg}$ for the full TDE sample to $e=\num{0.28}$ and $i=\qty{60.7}{deg}$ without AT~2020nov. The main change is in the radial parameters, with the outer-radius distribution contracting to $\xi_2=\qty{2751(916:3037)}{R_g}$, while the inner-radius median remains similar at $\xi_1=\qty{1388(1130:291)}{R_g}$. The emissivity and local broadening shift upward to $q=\num{2.60(0.48:0.30)}$ and $\sigma=\qty{779(363:830)}{\km \per \s}$. The broad Gaussian width is also largely unchanged, with the median increasing from $\mathrm{FWHM}=\qty{10880(6759:6486)}{\km \per \s}$ for the full TDE sample to $\mathrm{FWHM}=\qty{11503(1323:6356)}{\km \per \s}$ without AT~2020nov. Thus, AT~2020nov primarily drives the large-radius tail of the TDE distribution, while the lower eccentricities, moderate inclinations, and broad non-disk line widths are present in the remaining events.

Figure~\ref{fig:tde-parameter-evolution} shows that the inferred disk parameters evolve in an event-dependent way, with no single monotonic track shared by the full TDE sample. AT~2018hyz, which provides the densest temporal sampling, remains relatively compact over most epochs and generally favors low-to-moderate eccentricity after the double-peaked H$\alpha$ structure has emerged. AT~2020nov occupies the largest radial scales throughout its sequence, consistent with the interpretation that the flare illuminated an extended pre-existing disk, while AT~2020zso and PTF09djl retain high eccentricities in the available epochs but have insufficient temporal coverage to establish a robust evolutionary trend. The fitted parameters therefore appear to trace the instantaneous line-forming region, with no evidence for a universal circularization sequence across these events.

The inferred TDE radial scales are broadly consistent with simulation results in which circularization and reprocessing occur well outside the innermost disk. Hydrodynamic and magnetohydrodynamic calculations find that returning debris circularizes through stream self-intersection, shocks, and stream--disk interaction, producing eccentric, thick, and often wind-launching flows over hundreds to thousands of gravitational radii \citep{shiokawa2015,piran2015,krolik2016,bonnerot2020,steinberg2024,huang2024}. The compact radii inferred for AT~2018hyz and several other epochs are consistent with Balmer emission from this circularization or reprocessing region. These radii, however, are much larger than the scale of the innermost accretion flow. The absence of a universal monotonic trend in radius or eccentricity cautions against mapping the fitted line-profile parameters onto a single simulation coordinate, such as the mass-density peak or the circularization radius.

This interpretation is consistent with the radiative-transfer calculations of \citet{thomsen2026}, who model optical TDE spectra with optically thick, outflowing envelopes and find that the spectroscopic diversity of TDEs is largely governed by the ionization state of the gas, which is set by the balance between injected luminosity and envelope mass. In their framework, the extreme optical line widths arise primarily from electron scattering in the outflow, and the broad spectroscopic class of an event can remain relatively stable if luminosity and envelope mass decline together. Our fits do not model this reprocessing physics explicitly; the elliptical disk component instead captures the projected velocity asymmetry responsible for double-peaked or disk-like Balmer structure. However, the frequent need for an additional broad Gaussian component, with FWHM values of order $10^4\,\mathrm{km\ s^{-1}}$, is consistent with a substantial contribution from non-disk material such as an optically thick wind or reprocessing layer. In this sense, the disk parameters inferred here should be interpreted as describing the kinematic disk-like component of the line profile, while the broader Gaussian component likely absorbs part of the outflow- or scattering-dominated emission emphasized by \citet{thomsen2026}.

\begin{deluxetable*}{lccccccccc}
\tablecaption{Fitted disk parameters for TDE observations.\label{tab:tde_params_compact}}
\tablehead{
\colhead{TDE} & \colhead{Offset} & \colhead{$e$} & \colhead{$i$} & \colhead{$\xi_1$} & \colhead{$\xi_2$} & \colhead{$q$} & \colhead{$\sigma$} & \colhead{$\phi_0$} & \colhead{$\mathrm{FWHM}$} \\
\colhead{} & \colhead{[$\mathrm{days}$]} & \colhead{} & \colhead{[$\mathrm{deg}$]} & \colhead{[$\mathrm{R_g}$]} & \colhead{[$\mathrm{R_g}$]} & \colhead{} & \colhead{[$\unit{\km \per \s}$]} & \colhead{[$\mathrm{deg}$]} & \colhead{[$\unit{\km \per \s}$]}
}
\startdata
AT2018hyz & 10 & $0.28^{+0.24}_{-0.16}$ & $45.2^{+25.6}_{-17.1}$ & $207^{+246}_{-84}$ & $1741^{+2885}_{-1089}$ & $2.69^{+0.37}_{-0.54}$ & $740^{+763}_{-409}$ & $234.7^{+72.2}_{-71.3}$ & $11633^{+1924}_{-2474}$ \\
 & 28 & $0.21^{+0.19}_{-0.11}$ & $59.5^{+16.7}_{-21.1}$ & $852^{+411}_{-429}$ & $2328^{+4825}_{-1223}$ & $2.60^{+0.49}_{-0.52}$ & $740^{+1552}_{-437}$ & $132.8^{+74.6}_{-56.4}$ & $11261^{+915}_{-1205}$ \\
 & 40 & $0.54^{+0.15}_{-0.20}$ & $60.3^{+17.0}_{-21.0}$ & $1966^{+596}_{-856}$ & $3602^{+1513}_{-1594}$ & $2.84^{+0.40}_{-0.37}$ & $1025^{+486}_{-405}$ & $221.9^{+5.6}_{-11.2}$ & $11142^{+1382}_{-887}$ \\
 & 56 & $0.16^{+0.12}_{-0.04}$ & $58.8^{+17.6}_{-19.6}$ & $1514^{+503}_{-651}$ & $2697^{+883}_{-1166}$ & $2.37^{+0.34}_{-0.32}$ & $779^{+188}_{-203}$ & $259.4^{+34.8}_{-51.1}$ & $10486^{+1734}_{-1631}$ \\
 & 68 & $0.15^{+0.11}_{-0.03}$ & $65.4^{+14.0}_{-17.5}$ & $1664^{+380}_{-534}$ & $2782^{+597}_{-861}$ & $1.82^{+0.38}_{-0.37}$ & $614^{+149}_{-162}$ & $274.5^{+58.0}_{-36.5}$ & $10884^{+881}_{-832}$ \\
 & 80 & $0.12^{+0.08}_{-0.03}$ & $60.9^{+17.8}_{-19.8}$ & $1552^{+377}_{-661}$ & $2804^{+837}_{-1139}$ & $2.13^{+0.42}_{-0.42}$ & $410^{+168}_{-137}$ & $262.2^{+42.7}_{-55.4}$ & $10196^{+528}_{-531}$ \\
 & 98 & $0.16^{+0.13}_{-0.09}$ & $60.7^{+15.8}_{-16.0}$ & $1691^{+428}_{-566}$ & $2751^{+740}_{-910}$ & $2.31^{+0.48}_{-0.47}$ & $373^{+158}_{-118}$ & $217.1^{+9.7}_{-31.8}$ & $11503^{+454}_{-450}$ \\
 & 112 & $0.16^{+0.07}_{-0.04}$ & $66.6^{+13.0}_{-13.8}$ & $1433^{+351}_{-351}$ & $2733^{+573}_{-641}$ & $1.90^{+0.41}_{-0.40}$ & $436^{+186}_{-161}$ & $278.1^{+58.6}_{-30.4}$ & $10096^{+407}_{-401}$ \\
 & 132 & $0.20^{+0.08}_{-0.06}$ & $61.3^{+16.0}_{-17.6}$ & $1459^{+324}_{-530}$ & $2511^{+662}_{-913}$ & $2.34^{+0.43}_{-0.43}$ & $309^{+153}_{-81}$ & $241.7^{+9.8}_{-28.7}$ & $8785^{+356}_{-350}$ \\
 & 172 & $0.27^{+0.26}_{-0.15}$ & $58.5^{+18.1}_{-23.1}$ & $1308^{+803}_{-743}$ & $7695^{+18610}_{-4950}$ & $2.72^{+0.42}_{-0.51}$ & $1163^{+1202}_{-815}$ & $237.9^{+56.1}_{-58.6}$ & $10627^{+1517}_{-921}$ \\
AT2018zr & 12 & $0.37^{+0.20}_{-0.18}$ & $56.2^{+19.7}_{-21.7}$ & $410^{+637}_{-241}$ & $4526^{+10250}_{-3364}$ & $2.56^{+0.43}_{-0.42}$ & $1197^{+1302}_{-807}$ & $281.1^{+73.8}_{-67.8}$ & $14901^{+2686}_{-1257}$ \\
 & 26 & $0.41^{+0.21}_{-0.15}$ & $59.6^{+17.1}_{-20.2}$ & $1946^{+723}_{-930}$ & $4385^{+3306}_{-2043}$ & $2.96^{+0.43}_{-0.40}$ & $549^{+343}_{-249}$ & $115.2^{+23.2}_{-11.3}$ & $17322^{+1028}_{-896}$ \\
 & 40 & $0.28^{+0.20}_{-0.13}$ & $57.9^{+17.9}_{-20.5}$ & $1388^{+853}_{-913}$ & $5531^{+11643}_{-2901}$ & $2.54^{+0.48}_{-0.50}$ & $922^{+1001}_{-560}$ & $78.1^{+68.4}_{-71.0}$ & $23054^{+5891}_{-2376}$ \\
 & 52 & $0.50^{+0.08}_{-0.07}$ & $68.6^{+11.9}_{-15.1}$ & $509^{+127}_{-141}$ & $13314^{+4237}_{-3813}$ & $2.08^{+0.14}_{-0.12}$ & $1605^{+453}_{-300}$ & $98.7^{+23.9}_{-17.0}$ & $17204^{+4391}_{-3286}$ \\
 & 61 & $0.27^{+0.18}_{-0.12}$ & $61.5^{+15.9}_{-20.1}$ & $1677^{+853}_{-892}$ & $9152^{+19970}_{-5131}$ & $2.66^{+0.44}_{-0.51}$ & $1110^{+809}_{-673}$ & $103.8^{+61.7}_{-39.6}$ & $17626^{+2908}_{-1749}$ \\
AT2020nov & -24 & $0.24^{+0.11}_{-0.09}$ & $58.0^{+17.7}_{-18.2}$ & $568^{+314}_{-249}$ & $38421^{+11931}_{-16771}$ & $1.31^{+0.06}_{-0.09}$ & $338^{+38}_{-41}$ & $236.1^{+9.2}_{-29.6}$ & $3986^{+593}_{-323}$ \\
 & -9 & $0.31^{+0.04}_{-0.05}$ & $35.4^{+3.8}_{-2.6}$ & $253^{+48}_{-32}$ & $24564^{+4499}_{-3003}$ & $1.36^{+0.01}_{-0.01}$ & $400^{+7}_{-9}$ & $209.7^{+1.7}_{-3.8}$ & $--$ \\
 & 15 & $0.58^{+0.01}_{-0.01}$ & $29.6^{+1.2}_{-1.1}$ & $256^{+17}_{-15}$ & $17831^{+1311}_{-1139}$ & $1.55^{+0.01}_{-0.01}$ & $206^{+7}_{-4}$ & $202.9^{+0.7}_{-0.9}$ & $3746^{+133}_{-119}$ \\
 & 42 & $0.31^{+0.11}_{-0.11}$ & $55.4^{+19.3}_{-18.4}$ & $1030^{+508}_{-461}$ & $62369^{+24117}_{-28303}$ & $1.63^{+0.05}_{-0.05}$ & $366^{+41}_{-39}$ & $233.1^{+7.9}_{-22.8}$ & $--$ \\
 & 157 & $0.34^{+0.13}_{-0.10}$ & $59.7^{+16.8}_{-20.3}$ & $2901^{+2710}_{-1732}$ & $45333^{+29288}_{-21235}$ & $1.65^{+0.35}_{-0.28}$ & $345^{+170}_{-106}$ & $280.9^{+39.3}_{-28.9}$ & $3925^{+563}_{-272}$ \\
 & 164 & $0.25^{+0.06}_{-0.06}$ & $50.3^{+20.5}_{-12.1}$ & $958^{+456}_{-331}$ & $61037^{+30226}_{-21462}$ & $1.46^{+0.02}_{-0.03}$ & $390^{+35}_{-35}$ & $235.5^{+7.6}_{-14.2}$ & $3544^{+17}_{-8}$ \\
 & 268 & $0.17^{+0.23}_{-0.11}$ & $49.3^{+21.1}_{-18.0}$ & $6073^{+2507}_{-2988}$ & $87121^{+46806}_{-45346}$ & $1.74^{+0.35}_{-0.27}$ & $309^{+149}_{-76}$ & $224.6^{+65.0}_{-60.7}$ & $4150^{+937}_{-459}$ \\
 & 271 & $0.28^{+0.22}_{-0.16}$ & $48.7^{+23.8}_{-21.7}$ & $4307^{+3761}_{-3112}$ & $67228^{+84839}_{-44158}$ & $1.92^{+0.62}_{-0.49}$ & $501^{+409}_{-209}$ & $276.0^{+73.4}_{-77.6}$ & $4220^{+1066}_{-571}$ \\
AT2020zso & -3 & $0.84^{+0.08}_{-0.09}$ & $73.5^{+8.1}_{-10.5}$ & $478^{+75}_{-70}$ & $1847^{+322}_{-332}$ & $3.05^{+0.21}_{-0.18}$ & $416^{+422}_{-171}$ & $193.6^{+3.6}_{-4.9}$ & $22326^{+461}_{-483}$ \\
 & 10 & $0.93^{+0.02}_{-0.03}$ & $43.5^{+11.4}_{-3.8}$ & $259^{+53}_{-27}$ & $632^{+316}_{-98}$ & $2.90^{+0.26}_{-0.37}$ & $1631^{+359}_{-591}$ & $218.5^{+4.6}_{-8.4}$ & $4961^{+441}_{-806}$ \\
PTF09djl & 2 & $0.80^{+0.09}_{-0.14}$ & $67.6^{+10.5}_{-16.2}$ & $123^{+49}_{-19}$ & $693^{+519}_{-198}$ & $2.94^{+0.26}_{-0.24}$ & $3981^{+651}_{-1010}$ & $315.4^{+7.7}_{-12.7}$ & $19606^{+2190}_{-3158}$ \\
 & 31 & $0.91^{+0.03}_{-0.04}$ & $86.8^{+0.2}_{-0.3}$ & $103^{+3}_{-2}$ & $2043^{+173}_{-282}$ & $2.80^{+0.07}_{-0.07}$ & $2495^{+307}_{-300}$ & $330.8^{+2.0}_{-1.5}$ & $15849^{+782}_{-665}$ \\ \hline
All &  & $0.28^{+0.29}_{-0.11}$ & $59.5^{+6.9}_{-10.7}$ & $1308^{+597}_{-1051}$ & $4385^{+39842}_{-2296}$ & $2.34^{+0.50}_{-0.71}$ & $549^{+643}_{-200}$ & $234.7^{+43.1}_{-92.2}$ & $10880^{+6486}_{-6759}$ \\
\enddata
\tablecomments{The fitted epochs are based on spectra retrieved from WISeREP, except for AT~2020zso, whose reduced spectra were provided through private communication; see Section~\ref{sec:data}.}
\end{deluxetable*}


\section{Discussion} \label{sec:discussion}

\subsection{Comparing Disk Structure in TDEs and AGN}

\begin{figure*}
\centering
\includegraphics[width=\linewidth]{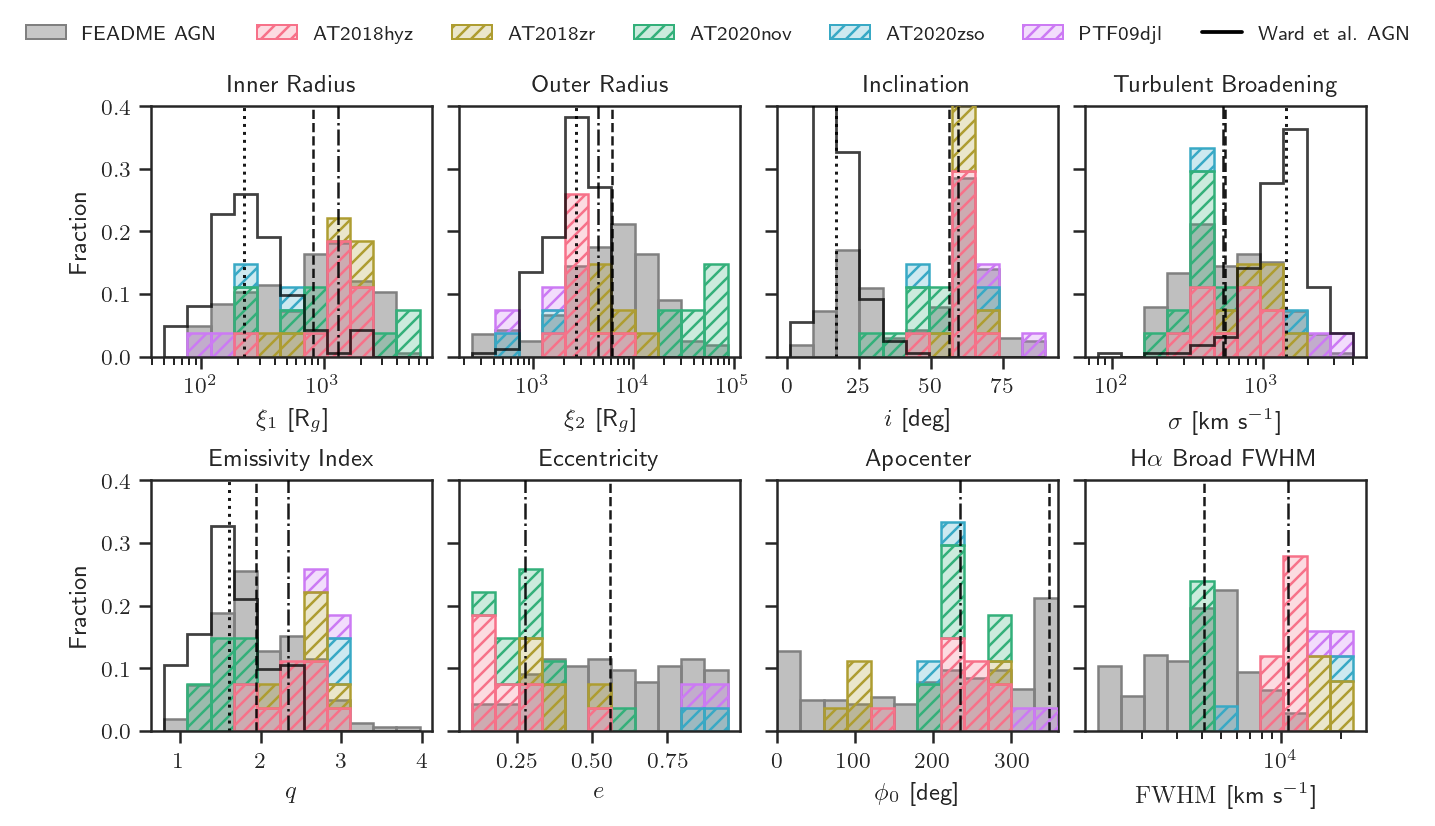}
\caption{\added{Empirical distributions of posterior-median parameters for the {\feadme} AGN and TDE fits, together with the published best-fitting circular-disk plus spiral-arm parameters from \citet{ward2024}. Grey histograms show the {\feadme} AGN, colored stacked histograms show the TDE epochs separated by event, and black step histograms show the Ward et al. values for the \num{162} AGN shared with the disk-bearing {\feadme} sample. The Ward et al. distributions are included for inclination, inner and outer radius, emissivity index, and local broadening, which are defined in both disk prescriptions. Black vertical lines mark the corresponding population medians: dashed lines represent the {\feadme} AGN, dash-dotted lines represent the {\feadme} TDEs, and dotted lines represent the Ward et al. AGN. The apocenter panel uses a circular median for the {\feadme} populations, and each population is normalized independently. The Ward et al. distributions use object-level best fits, whereas the {\feadme} distributions use posterior medians.}
\label{fig:tde-agn-parameter-distributions}}
\end{figure*}

\begin{deluxetable*}{lcccccc}
\tablecaption{\added{Shared disk parameters under the elliptical and circular plus spiral-arm prescriptions.}\label{tab:ward_tde_parameter_compare}}
\tablehead{
\colhead{Sample and model} & \colhead{$n$} & \colhead{$i$} & \colhead{$\xi_1$} & \colhead{$\xi_2$} & \colhead{$q$} & \colhead{$\sigma$} \\
\colhead{} & \colhead{} & \colhead{[$\mathrm{deg}$]} & \colhead{[$R_g$]} & \colhead{[$R_g$]} & \colhead{} & \colhead{[$\unit{\km \per \s}$]}
}
\startdata
{\feadme} elliptical AGN & 162 & $56.2^{+10.2}_{-35.8}$ & $812^{+1181}_{-607}$ & $6045^{+9067}_{-4226}$ & $1.94^{+0.64}_{-0.44}$ & $549^{+655}_{-252}$ \\
Ward et al. circular-disk plus spiral-arm AGN & 162 & $17.0^{+7.2}_{-6.0}$ & $225^{+270}_{-95}$ & $2685^{+1125}_{-1465}$ & $1.60^{+0.50}_{-0.40}$ & $1425^{+515}_{-545}$ \\
{\feadme} elliptical TDE & 27 & $59.5^{+6.9}_{-10.7}$ & $1308^{+597}_{-1051}$ & $4385^{+39842}_{-2296}$ & $2.34^{+0.50}_{-0.71}$ & $549^{+643}_{-200}$ \\
\enddata
\tablecomments{\added{Values are the medians and 16th--84th percentile ranges of the object- or epoch-level parameter estimates. The {\feadme} AGN row is restricted to the \num{162} objects with Ward et al. counterparts; the {\feadme} entries are posterior medians, and the Ward et al. entries are the published best-fitting values. The TDE row summarizes all \num{27} observed epochs.}}
\end{deluxetable*}

\added{
Figure~\ref{fig:tde-agn-parameter-distributions} and Table~\ref{tab:ward_tde_parameter_compare} compare the TDE parameters with two descriptions of the AGN profiles, but the two comparisons address different questions. Our primary comparison applies the {\feadme} elliptical model and the same inference procedure to both populations, retaining all \num{27} TDE epochs to represent the changing disk structure observed during each transient. This provides a uniform basis for comparing the projected line-emitting structures without introducing differences in the forward model or inference procedure. We use two-sample Kolmogorov--Smirnov tests to quantify the separation between the AGN and TDE parameter distributions, finding that the inner radius, outer radius, local broadening, and emissivity index distributions are not strongly separated ($p=0.44$, 0.28, 0.66, and 0.067, respectively), although the inclination distributions differ at $p=4.7\times10^{-3}$. The cluster-resolved comparison in Figure~\ref{fig:violin-parameter-compare} further shows that individual AGN morphology bins resemble the TDEs along different axes of disk structure.

The published circular-disk plus spiral-arm results of \citet{ward2024} provide a complementary test of model dependence by describing the same AGN spectra using a different representation of profile asymmetry. All five parameters shared by the Ward AGN and the {\feadme} TDE fits have different distributions ($p<10^{-4}$), with the Ward fits favoring lower inclinations, smaller radii, shallower emissivity profiles, and greater local broadening. Since the TDE spectra have not been fit with the Ward model, these results do not constitute a second controlled AGN--TDE comparison, but they show the scale and direction by which the physical interpretation of the AGN profiles changes when a spiral arm carries the asymmetry. The common-model comparison therefore establishes the relative placement of AGN and TDEs within the elliptical prescription, while the Ward comparison identifies which inferred similarities are sensitive to that prescription.

Within the common {\feadme} parameterization, the eccentricity distribution shows the clearest disk structure separation with the TDE epochs having a median $e=0.28$, below the AGN median of $e=0.56$, and the distributions differing at $p=1.6\times10^{-5}$. The lower TDE eccentricities may indicate that the line-emitting debris has undergone substantial circularization by the time broad Balmer emission emerges, consistent with hydrodynamic studies in which stream--disk collisions, pericenter shocks, or mixing at the nozzle reduce the orbital eccentricity of the returning debris \citep{steinberg2024,huang2024} and with the low eccentricities inferred for AT~2018hyz and PS18kh \citep[][respectively]{hung2020,holoien2019}. However, inefficient shocks or delayed inflow can preserve eccentric debris for longer periods, and the high-eccentricity epochs of AT~2020zso and PTF09djl may represent such cases, a more eccentric stream--disk configuration, or a preference induced by viewing geometry and limited spectral quality. In AGN, persistent profile asymmetries may instead arise from long-lived emissivity structure \citep{eracleous1995,strateva2003,storchi-bergmann2017}. Since the Ward model describes these asymmetries without orbital eccentricity and its spiral-arm parameters do not correlate significantly with the {\feadme} eccentricities, the measured $e$ contrast describes the greater projected asymmetry required by the AGN profiles within the elliptical model but does not establish a difference in physical orbital eccentricity.

Although the {\feadme} inclination distributions have similar medians ($i\approx \qty{56}{deg}$ for AGN and $i\approx\qty{59}{deg}$ for TDEs), their shapes differ at the $p=4.7\times10^{-3}$ level because the two lower-inclination morphology bins (C0 and C1) broaden the AGN distribution, whereas the TDEs are concentrated around moderate inclinations. The overlap between the TDEs and the higher-inclination AGN bins is consistent with the appreciable projected rotational velocities required to produce double-peaked profiles \citep{chen1989,eracleous2003}. This wider range may indicate that persistent AGN disks produce detectable disk-like Balmer structure over more viewing angles, while low-inclination TDE disks may be more readily obscured or diluted by reprocessing layers, winds, or other non-disk emission. The Ward solutions for the same AGN have a much lower median inclination of $i=\qty{17}{deg}$ and are strongly separated from the TDE distribution ($p<10^{-4}$), which under the circular plus spiral-arm prescription would place the TDE line-emitting regions at substantially higher inclinations than the AGN. However, the object-by-object shifts between the Ward and {\feadme} results show that inclination also compensates for the adopted radial scale, local broadening, and treatment of asymmetry. The similar inclinations inferred for the AGN and TDE samples indicate that their line profiles require comparable projected rotational velocities within the common elliptical disk model. However, the substantially lower AGN inclinations recovered by Ward show that this agreement does not establish that the physical viewing angle distributions of the two populations are intrinsically similar.

The low-inclination component of the {\feadme} distribution overlaps more closely with the Ward solutions, with the \num{61} matched AGN in C0 and C1 having a median {\feadme} inclination of approximately \qty{22}{deg}, close to the Ward median of \qty{17}{deg}, and their median {\feadme}-to-Ward inner-radius ratio is approximately unity. The other shared parameters show less agreement, and the inclinations remain uncorrelated object by object ($\rho_{\rm S}=0.03$, $p=0.83$), suggesting that these low-inclination solutions may occupy a region where the inferred inclination and inner radius are less sensitive to the asymmetry prescription. The available results do not establish that these objects form a distinct population of spiral-arm disks, nor can the higher-inclination {\feadme} component be associated with physically elliptical disks without a direct comparison between the competing models.

The emissivity index shows broad overlap in the common {\feadme} model, with a TDE median of $q\approx2.34$ compared with $q\approx1.94$ for the AGN and only marginal evidence for a difference between the distributions ($p=0.067$). Variations in optical depth, temperature gradients, illumination, and radiative transfer can all change the radial weighting of Balmer emission, and both disk-atmosphere and disk-wind calculations predict a broad range of effective emissivity slopes \citep{flohic2012,yong2017}, consistent with the absence of a universal value in AGN profile modeling \citep[e.g.][]{bon2009,kollatschny2013}. Within the elliptical prescription, the overlap indicates that the transient nature of the TDE disks does not require a systematically different radial concentration of Balmer emissivity once the disk-like profiles have emerged. In contrast, the Ward AGN have a shallower median of $q=1.60$ and are separated from the TDE distribution ($p<10^{-4}$), showing that part of the {\feadme} overlap arises from the way eccentricity and emissivity share the task of reproducing the profile asymmetry.

Local velocity broadening provides the closest agreement within the common {\feadme} model, with both populations having a median $\sigma\approx\qty{549}{\km \per \s}$ and no significant difference between their distributions ($p=0.66$). Because $\sigma$ represents unresolved motions and broadening within the line-emitting layer, this agreement suggests that similar effective velocity dispersions can accompany the Balmer-emitting structures in transient and persistent disks. The Ward AGN instead have a median $\sigma=\qty{1425}{\km \per \s}$, with their smaller radii and lower inclinations requiring greater local broadening to reproduce the observed line widths; the Ward and TDE distributions are significantly separated ($p<10^{-4}$). The same AGN profiles can therefore trade projected rotational broadening against local Gaussian broadening as the disk prescription changes, which limits the extent to which $\sigma$ can be interpreted directly as a common level of turbulence or vertical motion in the two populations.

The radial scales also overlap within the common elliptical model, with a TDE median inner radius of $\xi_1\approx1300\,R_g$ compared with approximately $820\,R_g$ for the AGN and no significant difference between the distributions ($p=0.44$). The outer-radius distributions are similarly consistent ($p=0.28$), with medians of approximately $4400\,R_g$ and $6000\,R_g$ for the TDEs and AGN, respectively. This overlap suggests that, once a Balmer-emitting disk-like structure is present, the observed line emission can arise over comparable gravitational scales despite the very different formation histories of a persistent AGN disk and newly assembled TDE debris. AT~2020nov is an important exception, reaching outer radii of several $10^4\,R_g$ that are consistent with a large preexisting disk illuminated by the flare \citep{earl2025}, while several other TDEs, including AT~2020zso and AT~2018hyz, favor compact line-emitting regions that may be associated with newly assembled disks near their circularization scales. The Ward AGN have smaller median inner and outer radii, $\xi_1=225\,R_g$ and $\xi_2=2685\,R_g$, and both radius distributions are separated from the TDE distributions ($p<10^{-4}$). The similar {\feadme} radii are therefore meaningful as effective line-forming scales under a common kinematic prescription, although the Ward comparison shows that they cannot be treated as model-independent physical boundaries of the AGN disks.

The apocenter angles ($\phi_0$) show a clearer difference when summarized with circular statistics, with a circular median near $\phi_0\simeq\qty{348}{deg}$ for the full AGN sample and $\phi_0\simeq\qty{235}{deg}$ for the TDEs. This shift in the preferred phase of the projected disk asymmetry provides a phenomenological description of the fitted line-profile shapes, although we treat apocenter separately from the linear KS comparisons because it is a circular coordinate. Since the apsidal orientation is not an intrinsic disk scale, depends on the projected viewing geometry and decomposition between disk and broad Gaussian emission, and can be weakly constrained when the disk eccentricity is modest, we regard this difference as secondary to the eccentricity, inclination, radial extent, and emissivity slope.

A further distinction between the two populations emerges in the FWHM of the additional broad H$\alpha$ component included when preferred by model selection. Although both AGN and TDEs frequently require this non-disk emission component, the TDEs have a median $\mathrm{FWHM}_{\rm TDE}\approx\qty{1.09e4}{\km \per \s}$, substantially broader than the AGN median of $\mathrm{FWHM}_{\rm AGN}\approx\qty{4.1e3}{\km \per \s}$, and the distributions are statistically distinct ($p<10^{-4}$). These values are consistent with optical TDE samples, which typically show hydrogen and helium lines with characteristic FWHM of order \qty{1e4}{\km \per \s} \citep{charalampopoulos2022,parkinson2022}, and with large AGN surveys, which generally find broad H$\alpha$ widths of a few $\sim$\qty{1e3}{\km \per \s} for classical broad-line regions \citep{shapovalova2009,negus2024}. The AGN broad Gaussian likely traces a more virialized BLR or disk-wind component, while the broader TDE feature may arise from shocks, reprocessing layers, or partially unbound debris in a dynamically younger environment. Unlike the absolute disk parameters, this comparison is made between components obtained with the same model and therefore provides more direct evidence that the non-disk line-emitting gas has a larger characteristic velocity in the TDEs.

Taken together, the common {\feadme} fits and the Ward comparison provide two complementary views of the disk structure. Within a uniform elliptical prescription, the AGN and TDE line-emitting regions occupy similar radial scales and require comparable emissivity slopes and local broadening, suggesting that disks with very different origins can produce Balmer emission from similar effective kinematic environments once the disk-like structure has formed. The lower fitted TDE eccentricities and substantially broader additional Gaussian components retain information about their different dynamical histories: the TDE profiles require less large-scale projected asymmetry from the disk component, while their non-disk emission arises at higher characteristic velocities. The Ward results show that the numerical agreement in inclination, radius, $q$, and $\sigma$ is not unique, because a circular disk with a spiral arm redistributes the same AGN profile structure among these parameters. Thus, the {\feadme} comparison identifies the relative differences that emerge when both populations are analyzed under the same assumptions, and the Ward comparison sets the systematic scale associated with interpreting those fitted parameters as the physical geometry of the disks.
}

\subsection{Limitations of the Disk Model} \label{subsec:disk-caveats}

\added{
The common elliptical-disk parameterization used here enables a controlled comparison between AGN and TDE line profiles, but describes all non-axisymmetric structure through nested, apsidally aligned eccentric orbits. For AGN DPEs, this model is degenerate with circular disks whose asymmetries are produced by azimuthally varying emissivity. Multi-epoch studies often favor evolving emissivity structures when the observed line-profile variability is compared with the long apsidal precession timescales of elliptical disks \citep{storchi-bergmann2003,gezari2007,lewis2010,eracleous2009,schimoia2017,ward2024,tucker2021}. Many analyses of AGN DPEs therefore commonly use circular or low-eccentricity disks with non-axisymmetric emissivity structure, although a fitted spiral arm may represent a spiral density wave, a hotspot, a fragmented emissivity enhancement, or another disk perturbation. Furthermore, agreement between a dynamical timescale and the observed variability does not prove that the fitted structure precesses or that it drives the optical continuum variability used to select the AGN sample, which may originate from instabilities at smaller radii than the Balmer-line-emitting disk. Consequently, the eccentricities measured here describe the projected asymmetry within the elliptical model and should not be interpreted as definitive orbital eccentricities for every AGN. Future implementations of additional non-axisymmetric disk components in {\feadme} will allow these interpretations to be compared directly.

The effect of this degeneracy can be measured using the results of \citet{ward2024}, who fit the same H$\alpha$ profiles with a circular disk carrying a spiral-arm emissivity perturbation. Published circular-disk solutions are available for \num{162} of the \num{165} disk-bearing AGN considered here, and approximately \num{94}\% have arm contrast ratios $A_{\rm s} \geq 2$. The parameter $A_{\rm s}$ measures the emissivity enhancement along the spiral arm relative to the underlying axisymmetric disk at the same radius, with larger values corresponding to greater arm--disk contrast. The Ward model therefore assigns a substantial azimuthal emissivity perturbation in nearly all cases. The shared disk parameters differ systematically between the two model families, wherein our elliptical fits have inner and outer radii larger by median factors of \num{3.5} and \num{2.4}, local broadening smaller by a median factor of \num{0.46}, inclinations higher by a median of \qty{34}{deg} (higher in \num{88}\% of objects), and emissivity indices steeper by \num{0.4}. These offsets are internally consistent because a more inclined and radially extended disk contributes greater projected rotational broadening and therefore requires less local broadening to reproduce the same overall line width. The inner radii retain a weak positive object-by-object correlation (Spearman $\rho_{\rm S}=0.23$), which may indicate that the inner emitting edge is more directly tied to the high-velocity profile structure, although its inferred scale remains strongly model-dependent. The remaining rank correlations are weak or absent ($\rho_{\rm S}=-0.06$, $0.16$, $0.07$, and $0.09$ for $\xi_2$, $\sigma$, $i$, and $q$, respectively), demonstrating that the differences cannot be described by a single rescaling of the elliptical parameters.

The fitted eccentricity shows no significant correlation with the \citet{ward2024} spiral-arm amplitude, width, or pitch angle ($\rho_{\rm S} \leq 0.08$ and $p > 0.3$), indicating that eccentricity is not a monotonic proxy for the fitted arm strength of individual objects and that the two model families describe the same broad profiles from different regions of parameter space. We cannot compare their fit quality formally because the published circular-disk results do not include the pointwise likelihoods needed for a common predictive score. ASASSN-14ko provides a useful example, as its asymmetric broad-line profile can be fit by both an elliptical disk and a circular disk with a spiral arm \citep{tucker2021}. Repeated passages of a partially disrupted or bound star could disturb the disk and produce this spiral structure, whose characteristic dynamical timescale of approximately \qty{140}{days} is close to the observed recurrence time of approximately \qty{114}{days}; relativistic precession of the corresponding elliptical disk would instead require \qtyrange{70}{700}{years}. This comparison favors periodically forced spiral structure in ASASSN-14ko, although the source is exceptional and the variability of ordinary AGN DPEs may have a different origin. Single-epoch profile fitting alone therefore cannot establish orbital eccentricity, and the absolute values of $\xi_1$, $\xi_2$, $i$, $q$, and $\sigma$ reported here remain conditional on the elliptical-disk prescription. Applying the same prescription to AGN and TDEs provides a controlled comparison in this model space, but Figure~\ref{fig:tde-agn-parameter-distributions} shows that the circular plus spiral-arm AGN solutions change the population comparison in every shared disk parameter.
}

A key assumption of the Eracleous elliptical disk model, and therefore of our {\feadme} fits, is that the line-emitting gas resides in a geometrically thin, optically thick accretion disk whose kinematics are dominated by nearly Keplerian orbital motion. This approximation follows the standard thin-disk picture \citep{shakura1973} and is most plausible for sub-Eddington AGN disks. Estimates for double-peaked broad-line AGN typically give Eddington ratios of $L/L_{\rm Edd}\sim10^{-3}$--$10^{-1}$ \citep{wu2004}. \citet{liu2017} interpret the eccentric disk fitted to PTF09djl as an optically thick, geometrically thin configuration at a sub-Eddington accretion rate. They also find that Balmer line emission can be produced in the illuminated surface layers of such disks.

For TDEs, however, this assumption is more tentative. A large fraction of optically selected TDEs appear to radiate near or above their Eddington limit at early times, implying that their accretion flows are likely to be geometrically thick and capable of launching powerful, optically thick winds \citep{strubbe2009,strubbe2011,metzger2016}. In AT~2018hyz, for example, \citet{hung2020} infer an Eddington ratio of $\sim 0.6$ at peak and argue that radiation pressure should puff up the disk and drive an optically thick outflow. Similar super-Eddington phases and associated reprocessing layers are generically expected in TDEs from theory and simulations, which predict quasi-spherical or funnel-shaped envelopes and strong disk winds that can obscure or reshape the underlying thin disk \citep{ohsuga2011,mckinney2015,sadowski2016,dai2018,bonnerot2020}. In this regime, the Balmer-emitting region may trace the photosphere of a thickened disk or the base of a wind, not necessarily the midplane of a thin disk.

The successful fits to the double-peaked Balmer lines in TDEs do not require the underlying flow to be a geometrically thin disk, since we use the Eracleous model as an effective parameterization of the projected velocity field in the line-forming region. Under this interpretation, the fitted inner and outer radii, emissivity index, and inclination encode where and how the H$\alpha$ photons are produced and escape, although the true geometry may involve a vertically extended disk atmosphere, a thick torus, or a combination of disk and outflow. This distinction is particularly relevant for events with high inferred Eddington ratios and strong evidence for reprocessing envelopes, in which our ``disk radii'' likely trace the characteristic orbital scales of the emitting material and may not correspond to sharp boundaries in a thin structure.

These caveats are less severe for the AGN sample, where the accretion flows may be closer to the classical thin-disk regime and where the same Eracleous formalism has been shown to reproduce the line profiles of long-lived double-peaked emitters \citep{eracleous1995,strateva2003}. Nevertheless, even in AGN there is strong evidence that disk winds and vertically extended broad-line regions can modify Balmer-line profiles and blur the distinction between thin-disk and wind-dominated geometries. For both AGN and TDEs, our {\feadme} parameters should therefore be interpreted as characterizing the kinematics and radial scales of the line-emitting gas under the thin-disk approximation, with the understanding that deviations from geometric thinness and the presence of optically thick outflows may systematically bias the mapping between these parameters and the true three-dimensional structure of the accretion flow.

\section{Conclusions} \label{sec:conclusions}

In this work we introduce {\feadme}, a new GPU-accelerated Python framework for modeling broad emission-line profiles using the relativistic elliptical accretion disk model of \citet{eracleous1995}. By coupling a fully differentiable forward model with modern Bayesian inference tools provided by {\jax} and {\numpyro}, {\feadme} enables efficient, large-sample modeling of disk-dominated Balmer-line emission in both AGN and TDEs. The flexibility and performance of this framework make it possible to conduct statistically rigorous, population-level comparisons of accretion-disk structure with uniform methodology.

We applied {\feadme} to 237 AGN from the DPE sample of \citet{ward2024} and to a sample of five TDEs known to exhibit disk-like Balmer emission. For every object, we evaluated three physically motivated model families and selected the optimal one using WAIC. The resulting posterior samples allowed us to reconstruct the distributions of disk geometry, kinematics, emissivity structure, and local broadening across both populations.

Our main findings are as follows:

\begin{enumerate}
    \item The AGN sample shows a diverse range of eccentricity, inclination, emissivity slope, radial extent, and local broadening. AGN disks populate a continuous manifold that can be usefully summarized with five phenomenological Gaussian-mixture morphology bins, without requiring sharply separated natural classes. This complexity likely reflects a combination of secular disk evolution, non-axisymmetric perturbations, and diverse radiative environments.

    \item \added{Within the common elliptical parameterization, the TDE and AGN samples overlap in radial scale, emissivity slope, and local broadening, while the TDEs show less large-scale projected asymmetry. However, comparison with the \citet{ward2024} circular-disk plus spiral-arm AGN solutions changes the inferred distributions, with the Ward AGN having substantially lower inclinations and radii, shallower emissivity profiles, and greater local broadening than the TDEs. The apparent overlap in these disk parameters is therefore model-dependent, and the fitted eccentricity contrast does not by itself measure a difference in physical orbital eccentricity.}

    \item Model selection indicates that a combination of disk emission and a broad Gaussian component provides the best description for the majority of both populations, although the kinematic properties of this broad component differ between the two samples. The systematically larger FWHM values found for the TDEs suggest that their non-disk emission arises from dynamically young, shock- or outflow-dominated gas, while the AGN component is more consistent with a virialized broad-line region.
\end{enumerate}

\added{Taken together, these results show that broad Balmer emission in both AGN and TDEs can contain a substantial rotating-disk contribution, although the inferred physical parameters depend on the model used to describe the profile asymmetry. Since elliptical disks and circular disks with non-axisymmetric emissivity can reproduce similar profiles while assigning different inclinations, radii, emissivity slopes, and local broadening, population comparisons must account for this model dependence.}

{\feadme} provides a scalable platform for future spectroscopic studies of accretion disks across cosmic time, from AGN variability surveys to time-domain studies of nuclear transients. As larger samples of TDEs and disk-emitting AGN become available from ZTF, LSST, and forthcoming spectroscopic follow-up programs, the ability to model disks with consistent physical assumptions will enable more robust analysis into the formation, evolution, and dynamics of accretion flows around supermassive black holes.

\begin{acknowledgments}

N.M.E., K.D.F., and M.S. acknowledge support from NSF grant AST–2206164. J.T.H acknowledges support provided by NASA through the NASA Hubble Fellowship grant HST-HF2-51577.001-A awarded by the Space Telescope Science Institute, which is operated by the Association of Universities for Research in Astronomy, Incorporated, under NASA contract NAS5-26555. M.E.V. acknowledges support from NSF grant AST-2307375. S.M. acknowledges support from grants NASA ADAP 80NSSC24K0666 and NASA NuSTAR data analysis funding 1729326.

\end{acknowledgments}

\appendix

\section{{\feadme} Interface and Template System} \label{appendix:cli}

To facilitate reproducibility and large-scale fitting, {\feadme} provides a command-line interface (CLI) with dedicated subcommands for different inference backends. All workflows share two primary inputs: (1) a spectral data file containing wavelength, flux, and flux uncertainty columns, and (2) a JSON template file that defines the physical model, parameter priors, and masking configuration.

The spectral data are supplied as an ASCII CSV file with three columns \texttt{wave} (in \unit{\angstrom}), \texttt{flux} (in \unit{mJy}), and \texttt{flux\_err} (in \unit{mJy}), which are read in and stored in a \texttt{Data} object. The observed-frame wavelength grid is used directly, with an optional internal rebinning step available for coarse-velocity binning. The accompanying template file is parsed into a \texttt{Template} object, which specifies the redshift, fractional white-noise hyperparameter, masking intervals, and the full set of disk and line profiles to be modeled.

\subsection{CLI Entry Points and Sampler Configuration}

The top-level CLI group \texttt{feadme} exposes a \texttt{run} subcommand that fits spectral data using the requested sampler and initializer configuration:

\begin{lstlisting}[language=bash, caption=CLI example for the {\feadme} NUTS workflow used in this work]
feadme run \
  --sampler=nuts \
  --template-path templates/ztf18aaaotwe.json \
  --data-path data/ZTF18aaaotwe.csv \
  --output-path results/ZTF18aaaotwe \
  --num-warmup=2000 \
  --num-samples=2000 \
  --num-chains=2 \
  --target-accept-prob=0.85 \
  --max-tree-depth=10 \
  --dense-mass \
  --integrator=mixed \
  --init-method=jax-lsq \
  --jax-lsq-start-method=structured \
  --jax-lsq-init-candidates=8 \
  --jax-lsq-init-steps=25 \
  --jax-lsq-batch-size=4 \
  --jax-lsq-selection-score=penalized-posterior
\end{lstlisting}

This command loads the template and data, performs a structured batched JAX least-squares initialization, constructs the {\numpyro} model, and runs NUTS with the requested configuration. Results are written to \texttt{results.nc} in the specified output directory, together with summary tables and diagnostic plots. If a \texttt{results.nc} file already exists and \texttt{--skip-existing} is enabled, {\feadme} will load the existing posterior instead of re-running the sampler.

In addition to standard NUTS, {\feadme} exposes a Neural Transport \citep[NeuTra;][]{hoffman2019} reparameterization backend through \texttt{--sampler=neutra}. NeuTra first fits an automatic variational guide to approximate the posterior geometry and then runs NUTS in the latent space defined by this transport map, which can improve sampling efficiency for posteriors with strong correlations or anisotropic curvature. Because its performance depends on the quality of the variational approximation, we treat NeuTra as an experimental backend and use standard NUTS for the production population analysis presented here.

The same \texttt{run} command also exposes alternative initialization schemes, including SVI, Pathfinder, MAP, and AutoDelta-style MAP initializers, through the \texttt{--init-method} option. These alternatives are useful for experimentation, but the production analysis in this paper uses \texttt{jax-lsq} initialization followed by NUTS sampling.

\subsection{Posterior Diagnostic Examples} \label{appendix:posterior-diagnostics}

\begin{figure*}
\centering
\includegraphics[width=\textwidth]{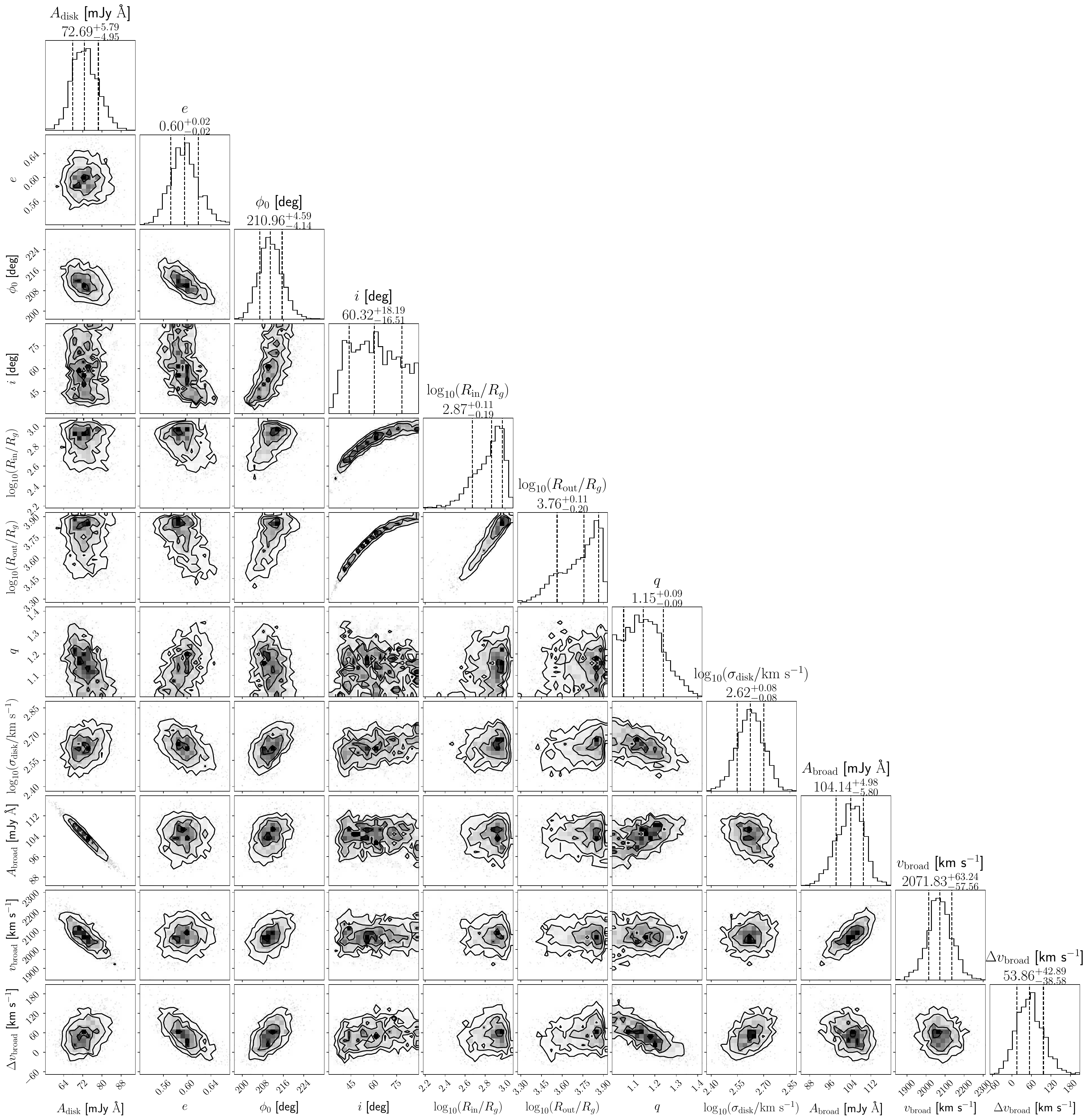}
\caption{Reduced posterior corner plot for the representative AGN Full Model fit to ZTF18aacdpbi. The corner plot shows the fitted disk parameters and the additional broad H$\alpha$ Gaussian component, omitting narrow-line nuisance parameters for clarity. Inner and outer radii are shown as $\log_{10}(R/R_g)$, the angular parameters are shown in degrees, and the area parameters are integrated component fluxes in \unit{mJy\angstrom}. The strongest correlations involve the disk and broad-component fluxes and the inclination--radius tradeoff characteristic of the AGN disk+broad+narrow fits.}
\label{fig:agn-corner-example}
\end{figure*}

\begin{figure*}
\centering
\includegraphics[width=\textwidth]{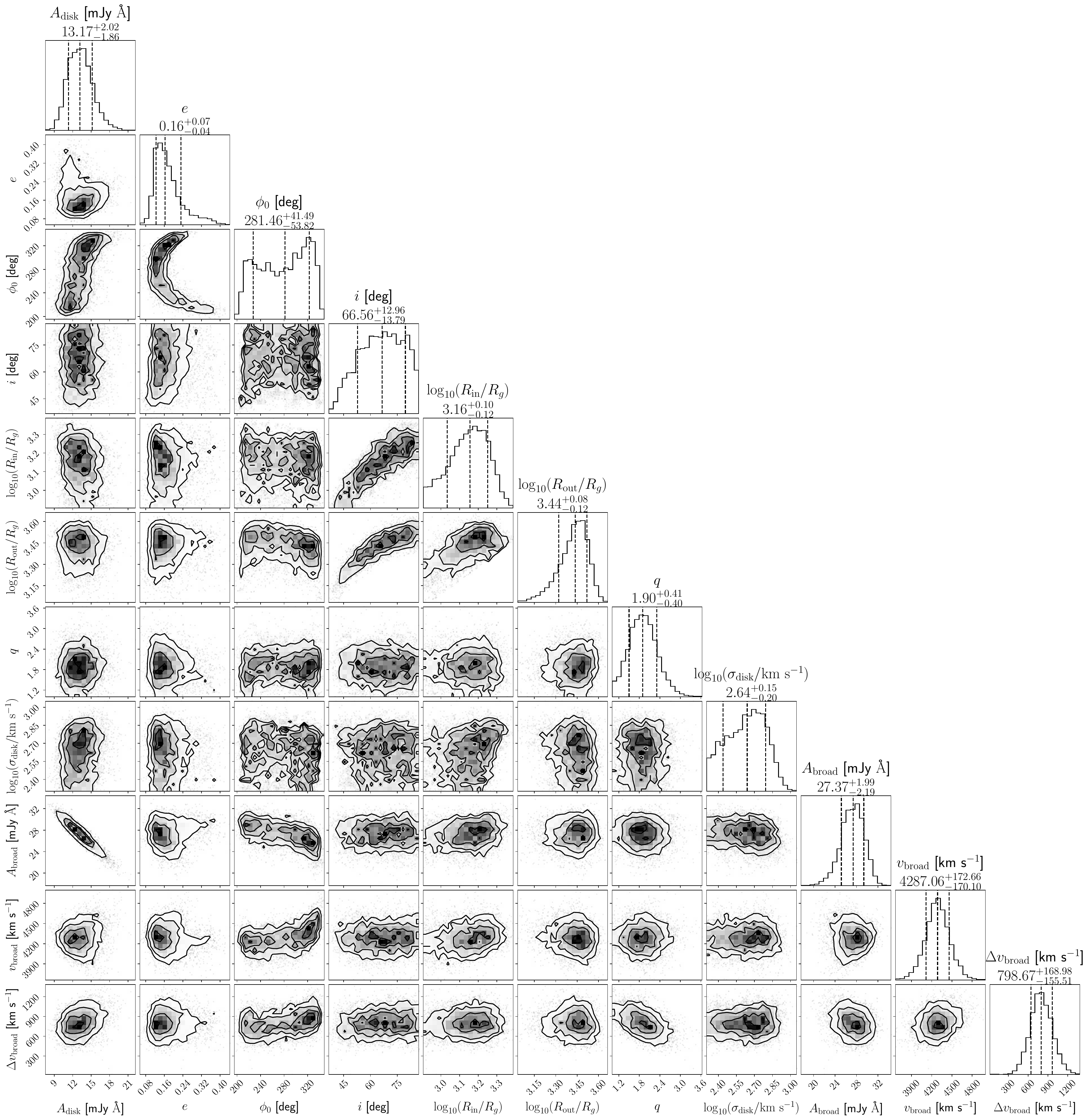}
\caption{Reduced posterior corner plot for the representative TDE Full Model fit to the AT~2018hyz spectrum observed at MJD 58539.5. The plotted parameter set matches Figure~\ref{fig:agn-corner-example}, allowing the AGN and TDE posterior geometries to be compared directly. As in the AGN example, the dominant correlations reflect the finite information content of a single broad-line profile and are incorporated into the reported posterior uncertainties.}
\label{fig:tde-corner-example}
\end{figure*}

Figures~\ref{fig:agn-corner-example} and \ref{fig:tde-corner-example} show reduced posterior corner plots for representative AGN and TDE fits drawn from the production sample, with the displayed parameters restricted to the disk structure and broad Gaussian component so that the main covariances remain visible. Both fits are retained in the analysis sample, and the AGN example falls in the gold quality tier and therefore has a comparatively clean posterior. Although these examples are not intended to represent typical members of either sample, they illustrate the covariance structure characteristic of single-epoch decompositions of broad Balmer profiles, including the tradeoff between disk and broad Gaussian flux, the coupling between disk inclination and radial scale, and the covariance between eccentricity and apsidal orientation.

\subsection{Template Structure}

The JSON template defines the physical and statistical structure of the model in a fully declarative way. At the top level, a template contains:

\begin{itemize}
    \item \texttt{"name"}: an identifier for the object;
    \item \texttt{"obs\_date"}: an optional observation timestamp (used for time-series analyses);
    \item \texttt{"redshift"}: a parameter block describing the redshift prior;
    \item \texttt{"log\_frac\_noise"}: a parameter block for the fractional additional white-noise term;
    \item \texttt{"disk\_profiles"}: a list of elliptical disk components (typically one per broad Balmer line);
    \item \texttt{"line\_profiles"}: a list of Gaussian line components (narrow and broad);
    \item \texttt{"mask"}: a list of wavelength intervals to include in the fit.
\end{itemize}

Each parameter block (for example, \texttt{"redshift"} or \texttt{"sigma"} in the disk profile) is itself a small dictionary with a common schema:
\texttt{distribution} (e.g., \texttt{"uniform"}, \texttt{"log\_uniform"}, \texttt{"normal"}, etc.), \texttt{low}, \texttt{high}, \texttt{loc}, \texttt{scale}, a Boolean \texttt{fixed} flag, and an optional \texttt{shared} field that links parameters across components. The \texttt{circular} flag indicates angular parameters (such as the apocenter angle) for which circular statistics are used internally.

The example below shows a shortened template used in this work to fit a double-peaked H$\alpha$ profile with an elliptical disk plus narrow and broad Gaussian components, as well as the neighboring \elinedbl[f]{N}{ii}{6548}{6584} and \elinedbl[f]{S}{ii}{6716}{6731} narrow lines:

\begin{lstlisting}[language=json, caption=Excerpt of a {\feadme} template for ZTF18aaaotwe. In-depth descriptions of parameter definitions can be found on the GitHub repository\footref{fn:gh-url}.]
{
  "name": "ZTF18aaaotwe",
  "obs_date": 0.0,
  "redshift": {
    "distribution": "uniform",
    "fixed": false,
    "low": 0.06533,
    "high": 0.06686
  },
  "log_frac_noise": {
    "distribution": "uniform",
    "fixed": true,
    "value": -10.0
  },

  "disk_profiles": [
    {
      "name": "halpha_disk",
      "center": 6562.819,
      "area": {
        "distribution": "uniform",
        "fixed": false,
        "low": 0.0,
        "high": 49.8
      },
      "offset": {
        "distribution": "normal",
        "fixed": true,
        "value": 0.0
      },
      "inner_radius": {
        "distribution": "log_uniform",
        "fixed": false,
        "low": 100.0,
        "high": 5000.0
      },
      "radius_ratio": {
        "distribution": "log_uniform",
        "fixed": false,
        "low": 1.2,
        "high": 22.0
      },
      "eccentricity": {
        "distribution": "normal",
        "fixed": false,
        "low": 0.0,
        "high": 0.95,
        "loc": 0.35,
        "scale": 0.25
      },
      "inclination": {
        "distribution": "beta",
        "fixed": false,
        "alpha": 1.5,
        "beta": 1.5,
        "low": 0.0,
        "high": 1.5698
      },
      "apocenter": {
        "distribution": "uniform",
        "fixed": false,
        "circular": true,
        "low": 0.0,
        "high": 6.2832
      },
      "q": {
        "distribution": "normal",
        "fixed": false,
        "low": 1.0,
        "high": 4.0,
        "loc": 2.5,
        "scale": 0.5
      },
      "sigma": {
        "distribution": "log_uniform",
        "fixed": false,
        "low": 200.0,
        "high": 3000.0
      },
      "baseline": {
        "distribution": "uniform",
        "fixed": true,
        "value": 0.0
      }
    }
  ],

  "line_profiles": [
    {
      "name": "halpha_narrow",
      "shape": "gaussian",
      "center": 6562.819,
      "area": {
        "distribution": "uniform",
        "fixed": false,
        "low": 0.0,
        "high": 32.4
      },
      "offset": {
        "distribution": "normal",
        "fixed": true,
        "value": 0.0
      },
      "vel_width": {
        "distribution": "normal",
        "fixed": false,
        "low": 70.0,
        "high": 500.0,
        "loc": 162.5,
        "scale": 25.0
      }
    },

    {
      "name": "halpha_broad",
      "shape": "gaussian",
      "center": 6562.819,
      "area": {
        "distribution": "uniform",
        "fixed": false,
        "low": 0.0,
        "high": 49.8
      },
      "offset": {
        "distribution": "normal",
        "fixed": false,
        "low": -2000.0,
        "high": 2000.0,
        "loc": 0.0,
        "scale": 400.0
      },
      "vel_width": {
        "distribution": "log_normal",
        "fixed": false,
        "low": 500.0,
        "high": 6000.0,
        "loc": 2200.0,
        "scale": 900.0
      }
    },

    // Additional [N II] and [S II] narrow components omitted for brevity.
    // Their rest-frame centers are fixed scalars, while velocity widths
    // are shared with halpha_narrow through the vel_width.shared field.
  ],

  "mask": [
    { "lower_limit": 6824.086, "upper_limit": 7244.727 }
  ]
}
\end{lstlisting}

In this example, the H$\alpha$ disk profile is defined by a fixed rest-frame line center, a fitted velocity offset, an integrated line area, and the elliptical disk parameters used throughout this work (inclination, inner radius, radius ratio, eccentricity, apocenter angle, emissivity index $q$, and local broadening $\sigma$). The outer radius is not specified independently in the template, but is derived deterministically as $\xi_2 = \xi_1 \times$ \texttt{radius\_ratio}. Gaussian line components follow the same convention: \texttt{center} is the fixed rest-frame wavelength of the transition, while \texttt{offset} encodes any velocity displacement from that center. Several of the forbidden narrow lines share the same velocity width as the narrow H$\alpha$ component through the \texttt{shared} field.

Because the modeling logic is fully encoded in the template, users can construct a wide range of models (pure disks, disk plus broad-line components, or line-only models) without modifying the source code. The same template description is used for both NUTS and SVI runs, enabling consistent comparisons across sampler types and facilitating survey-scale, reproducible modeling of double-peaked emitters.

The full {\feadme} source code, including the CLI, sampler implementations, template system, and all supporting modules, is preserved in the version 0.2.1 Zenodo archive \citep{earl_2026_22869620}. Ongoing development is publicly available on GitHub\footref{fn:gh-url}. The project welcomes community engagement: users are encouraged to submit issues, request features, open discussions, or contribute enhancements through pull requests. By maintaining {\feadme} as an open and extensible platform, our goal is to support reproducible analysis workflows and facilitate broader community involvement in the study of disk-emitting AGN and TDEs.

\bibliography{references}{}
\bibliographystyle{aasjournalv7}

\end{document}